\documentclass[twocolumn]{IEEEtran}
\usepackage{caption}
\usepackage[T1]{fontenc}
\usepackage[latin1]{inputenc}
\usepackage[english]{babel}
\usepackage{amsmath}
\usepackage{amsthm}
\usepackage{dsfont}
\usepackage{amsfonts}
\usepackage{mathrsfs}
\usepackage{pifont}
\usepackage{amssymb}
\usepackage{verbatim}
\usepackage{booktabs}
\usepackage{upgreek}
\usepackage[final]{graphicx}
\usepackage{subfigure}
\usepackage{epsfig}
\usepackage{bm}
\usepackage[nolist,nohyperlinks]{acronym}
\usepackage{float} 
\usepackage{color}
\usepackage[table]{xcolor}
\usepackage{multirow}
\usepackage{fancyhdr}
\usepackage[colorlinks]{hyperref}
\usepackage{cite}
\usepackage{balance}
\usepackage{algorithm}
\usepackage{algorithmic}
\usepackage[nolist,nohyperlinks]{acronym}
\usepackage{url}
\usepackage{orcidlink}
\usepackage{rotating}

\newcommand{\real}{\text{Re}}
\newcommand{\imag}{\text{Im}}
\newcommand{\bzero}{\mbox{\boldmath{$0$}}}

\newcommand{\bA}{\mbox{\boldmath{$A$}}}
\newcommand{\ba}{\mbox{\boldmath{$a$}}}

\newcommand{\bC}{\mbox{\boldmath{$C$}}}

\newcommand{\bE}{\mbox{\boldmath{$E$}}}

\newcommand{\bG}{\mbox{\boldmath{$G$}}}

\newcommand{\bH}{\mbox{\boldmath{$H$}}}

\newcommand{\bK}{\mbox{\boldmath{$K$}}}
\newcommand{\bI}{\mbox{\boldmath{$I$}}}

\newcommand{\bJ}{\mbox{\boldmath{$J$}}}

\newcommand{\bS}{\mbox{\boldmath{$S$}}}

\newcommand{\bT}{\mbox{\boldmath{$T$}}}

\newcommand{\bU}{\mbox{\boldmath{$U$}}}

\newcommand{\bV}{\mbox{\boldmath{$V$}}}

\newcommand{\bX}{\mbox{\boldmath{$X$}}}
\newcommand{\bx}{\mbox{\boldmath{$x$}}}
\newcommand{\bY}{\mbox{\boldmath{$Y$}}}
\newcommand{\by}{\mbox{\boldmath{$y$}}}

\newcommand{\balpha}{\mbox{\boldmath{$\alpha$}}}

\newcommand{\tr}{\mbox{\rm tr}\, }

\newcommand{\diag}{\mbox{\boldmath\bf diag}\, }

\def\E{{\mathbb E}}

\newtheorem{remark}{Remark}

\begin{acronym}
\acro{GML}{generalized maximum likelihood}
\acro{i.i.d.}{independent and identically distributed}
\acro{MIMO}{multiple-input multiple-output}
\acro{ML}{maximum likelihood}
\acro{MOS}{model order selection}
\acro{pdf}{probability density function}
\acro{PolSAR}{polarimetric synthetic aperture radar}
\acro{psd}{positive semi-definite}
\acro{RMSE}{root mean square error}
\acro{SCM}{sample covariance matrix}
\acro{SNR}{signal-to-noise ratio}
\end{acronym}

\begin{document}
\title{Covariance Symmetries Classification in Multitemporal/Multipass PolSAR Images}

\author{Dehbia Hanis\orcidlink{0009-0009-4173-8806}, Luca Pallotta\orcidlink{0000-0002-6918-0383},~\IEEEmembership{Senior Member,~IEEE}, Augusto Aubry\orcidlink{0000-0002-5353-0481},~\IEEEmembership{Senior Member, IEEE},\\ Aichouche Belhadj-Aissa\orcidlink{0000-0001-7287-2981}, and Antonio De Maio\orcidlink{0000-0001-8421-3318},~\IEEEmembership{Fellow, IEEE}
\thanks{Manuscript received September 13, 2024. This research activity of Dehbia Hanis has been conducted during her visit at University of Basilicata, under the local scientific supervision of Dr. Luca Pallotta.\\
({\it Corresponding author: Antonio De Maio}.)}
\thanks{D. Hanis, and A. Belhadj-Aissa are with the Laboratory of Image Processing and Radiation, University of Sciences and Technology Houari Boumediene (USTHB), Algiers, Algeria. E-mail: dehbia\textunderscore hanis@yahoo.com, houria.belhadjaissa@gmail.com.}
\thanks{L. Pallotta is with Department of Engineering (DiING), University of Basilicata, via dell'Ateneo Lucano 10, 85100 Potenza, Italy. E-mail: luca.pallotta@unibas.it.}
\thanks{A. Aubry and A. De Maio are with the Department of Electrical Engineering and Information Technology (DIETI), Universit\`{a} degli Studi di Napoli ``Federico II'', via Claudio 21, I-80125 Napoli, Italy, and with National Inter-University Consortium for Telecommunications (CNIT), Parma, Italy. E-mail: \{augusto.aubry, ademaio\}@unina.it.}
}

\markboth{IEEE Trans. on Geoscience and Remote Sensing}{Hanis et al.}

\maketitle

\begin{abstract}
A polarimetric synthetic aperture radar (PolSAR) system, which uses multiple images acquired with different polarizations in both transmission and reception, has the potential to improve the description and interpretation of the observed scene. This is typically achieved by exploiting the polarimetric covariance or coherence matrix associated with each pixel, which is processed to meet a specific goal in Earth observation. This paper presents a design framework for selecting the structure of the polarimetric covariance matrix that accurately reflects the symmetry associated with the analyzed pixels. The proposed methodology leverages both polarimetric and temporal information from multipass PolSAR images to enhance the retrieval of information from the acquired data. To accomplish this, it is assumed that the covariance matrix (of the overall acquired data) is given as  the Kronecker product of the temporal and polarimetric covariances. An alternating maximization algorithm, known as the flip-flop method, is then developed to estimate both matrices while enforcing the symmetry constraint on the polarimetric covariance. Subsequently, the symmetry structure classification is formulated as a multiple hypothesis testing problem, which is solved using model order selection techniques. The proposed approach is quantitatively assessed on simulated data, showing its advantages over its competitor, which does not exploit temporal correlations. For example, it reaches accuracies of $94.6\%$ and $92.0\%$ for the reflection and azimuth symmetry classes, respectively, while the competitor achieves $72.5\%$ and $72.6\%$ under the same simulation conditions. Moreover, the proposed method can realize a Cohen's kappa coefficient of $0.95$, which significantly exceeds that of its counterpart equal to $0.78$. Finally, the effectiveness of the proposed framework is further demonstrated using measured RADARSAT-2 data, corroborating the results obtained from the simulations. Specifically, tests conducted applying the Freeman-Durden Wishart classification have proved that the new approach greatly enhances the accuracy of pixel classification. For instance, in areas dominated by surface scattering, it boosts the percentage of correctly classified pixels from $68.23\%$, achieved using the classic method, to $91.65\%$.
\end{abstract}


\section{Introduction}

\IEEEPARstart{M}{odern} \ac{PolSAR} systems that use multiple images acquired with distinct polarizations in both transmission and reception have the potential to improve the description and interpretation of the observed scene (including enhancements in the classification, detection and/or recognition capabilities of the entire system) \cite{aubry2022polarimetric}. This process typically involves statistical inference from the data corresponding to individual pixels that can gathered via appropriate probing schemes \cite{aubry2022polarimetric, guodong2022quasi}. In this context, the polarimetric covariance (or coherence) matrix is one of the primary tools used to infer the scattering mechanisms that characterize the observed objects of the analyzed scenario. The polarimetric covariance matrix contains hidden information that can help differentiating various aspects in the SAR image. This includes distinguishing between different areas, such as land versus sea, bare fields versus cultivated ones, or buildings versus trees. It also allows for the discrimination between categories within the same type, like small-stem crops versus broad-leaf crops, as well as identifying different scattering mechanisms, such as dipole, diplane, and trihedral. In this respect, several polarimetric decompositions based on coherence matrix \cite{FreemanDurdenDecomp98, Yamaguchi2005, Yamaguchi2011, Cloude1997, PottierBook} aid in identifying different scattering mechanisms like surface, volume, and double-bounce scattering. The interested reader may also refer to \cite{ainsworth2022model, eltoft2018polarimetric, eltoft2019model} for further approaches. Additionally, several strategies for scene classification/segmentation (typically categorized into unsupervised or supervised algorithms) exploiting the polarimetric information in SAR images can be found, for instance, in \cite{VanZyl1989, lemoine1994polarimetric, 964969, 1526759, 789621, ParkMoonPottier, dabboor2013new, dimartino2019closed, 1288367, 10098835, wang2024geometry, 10463089}. The unsupervised method involves grouping image pixels based on shared characteristics or features, and operates automatically without user intervention. On the other hand, the supervised method necessitates user involvement, where algorithms are trained using pre-classified reference samples to classify new, unknown data. For instance, in \cite{VanZyl1989}, both amplitude and phase information from the polarimetric channels are used to distinguish different scattering behaviors, allowing the interpretation of radar images in terms of surface characterization of various types of terrain. Analogously in \cite{ParkMoonPottier}, the authors propose a new model for vegetation scattering mechanisms in mountainous forests, extending the classic radiative transfer model by accounting for the sloping ground surface beneath the vegetation canopy.

Most of the above analyses rely on second-order moment-based metrics, namely the covariance and coherence matrices, which are object of an estimation process. Often, the covariance matrix is obtained as the \ac{SCM} which is its \ac{ML} estimate assuming a homogeneous environment (i.e., zero-mean \ac{i.i.d.} circularly symmetric Gaussian data). This involves taking the outer product of the observation vector with its Hermitian transpose and then spatial averaging over a large enough window of pixels to get the overall \ac{SCM}. Unfortunately, the \ac{SCM} suffers from some limitations, such as the scarcity of homogeneous data that can be extracted from the spatial neighborhood of the pixel under test. As a consequence, before applying any of the above decompositions, more sophisticated covariance estimators can be used in place of the \ac{SCM} to exploit intrinsic characteristics of the observed scene. In particular, the polarimetric covariance matrix may exhibit a distinct structure when significant scattering symmetry features (i.e., reflection symmetry, rotation symmetry, and azimuth symmetry) dominate the radar return \cite{Nghiem1992,PottierBook}. Hence, it becomes crucial to establish the presence of specific symmetries as a preliminary stage for image processing interpretation. In fact, by examining these symmetries, it can be gained insight into the physical properties and nature (e.g., vegetation, water bodies, urban areas) of SAR backscatters in the observed region. Notably, enforcing symmetry in the estimation process allows for better exploitation of the underlying scattering mechanisms for classification purposes. Moreover, since fewer parameters need to be estimated, fewer looks are required, making the process less sensitive to spatial variability of the parameters. In such a context, in \cite{pallotta2016detecting, pallotta2022covariance}, a \ac{PolSAR} covariance symmetries detection framework has been proposed and analyzed. The problem has been framed as a multiple hypothesis testing task involving nested instances (the four hypotheses refer to absence of symmetry, reflection symmetry, rotation symmetry, and azimuth symmetry \cite{Nghiem1992, PottierBook}), and several \ac{MOS} criteria \cite{Selen2004, stoica2004model, stoica2004information, Kay2005} have been applied to classify each pixel according to its respective covariance structure. The core objective of the \ac{MOS} rules is to minimize the Kullback-Leibler divergence between the proposed set of candidate models and the actual data distribution \cite{Selen2004, stoica2004model, stoica2004information, Kay2005}. In fact, they address the limitations of the \ac{GML} method \cite{KayBook, KayLetter} by incorporating a penalty term, which accounts for the number of unknown parameters, into the likelihood function for each hypothesis prior to its maximization. This is necessary because the \ac{GML} fails completely in the presence of nested hypotheses, resulting in the selection of the instance that encompasses all others. It is also worth underlining that the polarimetric covariance structure classification framework in \cite{pallotta2016detecting} has been also extended to account for data heterogeneity \cite{pallotta2019robust}, when data are filtered before performing inferences \cite{pallotta2023screening, pallotta2023outlier}, as well as assuming that the pixels in the extracted neighborhood do not share the same symmetry \cite{han2023innovative}.

This paper for the first time introduces a methodology for the automatic classification of the symmetric structure of the polarimetric covariance matrix in multipass/multitemporal \ac{PolSAR} imagery. The approach aims to enhance the estimation of pixel polarimetric characteristics by processing multiple observations of the same scene while accounting for temporal correlation. This establishes a comprehensive framework that extends beyond the single-image symmetry detection method outlined in \cite{pallotta2016detecting}, by jointly processing all available multitemporal acquisitions, and hence exploiting the additional information arising from observing more than one image of the same scene. In particular, the data are modeled as Gaussian-distributed, where the covariance matrix is expressed as the Kronecker product of temporal and polarimetric covariance matrices. The polarimetric component is constrained to conform to the specific symmetry properties above described, while the temporal covariance is assumed to be positive definite but unstructured. In addressing the resulting problem, the estimate of the Kronecker-structured temporal/polarimetric covariance matrix is derived using a \emph{flip-flop} algorithm, which alternates between estimating the temporal and polarimetric covariances while enforcing the symmetry structures, and ensuring its convergence to a stationary point of the design optimization problem. Notice that, for the case of absence of symmetry (e.g., for both temporal and polarimetric unstructured matrices), the covariance estimate can be obtained as the solution of the flip-flop (or \emph{ping-pong}) algorithm in \cite{Werner2008}. Subsequently, to accurately select the polarimetric structure that best represents the dominant symmetry of the pixels' return, several \ac{MOS} criteria are synthesized. They integrate the \ac{ML} estimates of the covariance structures identified in the previous stage. Extensive simulations are conducted to quantitatively demonstrate the effectiveness of the developed strategy, also with respect to its competitor. Finally, the devised procedure is also validated on a pair of real-recorded \ac{PolSAR} images from RADARSAT-2. To facilitate the reader's understanding, it is important to disclose here that the innovations introduced by the proposed framework can be summarized in the following steps:
\begin{enumerate} 
\item introduction of a bespoke model for structured covariance matrix estimation in the presence of multiple polarimetric SAR acquisitions;
\item design of a \emph{flip-flop} based algorithm for the estimation of the (unstructured) temporal and (structured) polarimetric covariance matrices;
\item synthesis of \ac{MOS} rules for selecting the correct hypothesis in terms of covariance structure;
\item extensive simulations and tests on real-recorded data.
\end{enumerate}

The remainder of the paper is structured as follows: Section \ref{ProbFormulation} describes the process for multipass datacube formation, while Section \ref{sec_background} summarizes the covariance matrices under the different symmetries' hypothesis. In Section \ref{problem_formulation} the estimation and classification problems are formulated, and the solution for the Kronecker structured covariance estimate is developed. Then, Section \ref{sec_MOS} presents the proposed framework for polarimetric covariance symmetries classification. The performance of the devised technique applied on simulated and real X-band SAR data is presented and discussed in Section \ref{secPerformanceAnalysisAndDiscussions}. Finally, some concluding remarks are given in Section \ref{sec_conclusion}.

\section*{Notation}
Boldface is used for vectors $\ba$ (lower case), and matrices $\bA$ (upper case). The conjugate and conjugate transpose operators are denoted by the symbols $(\cdot)^*$ and $(\cdot)^H$ respectively, whereas the symbol $(\cdot)^{-1}$ denotes the inverse. $\tr\{\cdot\}$ and $|\cdot|$ are respectively the trace and the determinant of the square matrix argument. $\bI$ and $\bzero$ denote respectively the identity matrix and the matrix with zero entries (their size is determined from the context). The vector $\mathbf{vec\{\cdot \}}$ is obtained by stacking the columns of its matrix argument on top of each other. $\diag(\ba)$ indicates the diagonal matrix whose $i$-th diagonal element is the $i$-th entry of $\ba$. The letter $j$ represents the imaginary unit (i.e. $j=\sqrt{-1}$). For any complex number $x$, $|x|$ represents the modulus of $x$, $\real\{x\}$ is its real part, and $\imag\{x\}$ is its imaginary part. Moreover, ${\cal{H}}_{n}^{++}$ is the set of $n\times n$ \ac{psd} Hermitian matrices, ${\cal{S}}_{n}^{++}$ is the set of $n\times n$ psd symmetric matrices, and ${\cal{P}}_{n}^{++}$ is the set of $n\times n$ real psd persymmetric matrices. A real persymmetric matrix $\bC_n$ is a matrix with the following property $\bC_n = \bJ\bC_n^T\bJ,$ with $\bJ$ the $n\times n$ permutation matrix $$\bJ = \left[ \begin{matrix}
0 & 0 & \cdots & 0 & 1 \\
0 & 0 & \cdots & 1 & 0 \\
\vdots & \vdots &  & \vdots & \vdots \\
0 & 1 & \cdots & 0 & 0 \\
1 & 0 & \cdots & 0 & 0 \\
\end{matrix} \right].$$
The symbols $\otimes$ and $\odot$ indicate the Kronecker and the element-wise or Hadamard product, respectively. Finally, $\E\left[\cdot\right]$ denotes statistical expectation.

\section{Multipass datacube definition}\label{ProbFormulation}

A \ac{PolSAR} system collects $N = 4$ complex returns for each observed image pixel, leveraging four different polarimetric channels, viz., HH, HV, VH, and VV. Generally, the number of polarimetric returns is reduced to $N = 3$ since the reciprocity theorem is assumed valid \cite{aubry2019assessing, pallotta2020reciprocity} and the HV and VH channels are substituted by their coherent average \cite{polsarpro}, namely $\left(\text{HV} + \text{VH}\right)/2$, or a single cross-polarimetric channel is used. 

With reference to the case $N = 3$, the polarimetric returns regarding the same pixel are hence organized in the vector $\by = \left[y_{\text{HH}}, y_{\text{HV}}, y_{\text{VV}}\right]^T$, where $y_{t r}$ denotes the complex polarimetric return in the transmitting and receiving polarization $t \in \{\text{H}, \text{V}\}$ and $r\in \{\text{H}, \text{V}\}$, respectively. As a consequence, denoting by $L$ and  $C$ the vertical and horizontal size of the image, respectively, a datacube $\bY$, composed by the vectors $\bY(c, l)$, $l = 1,\ldots, L$, $c = 1,\ldots, C$, each of size $N$ is formed. Therefore, the sensor provides a 3-D data stack $\bY$ of size $L \times C \times N$ which is referred to as datacube in the following. 

\begin{figure}[ht!]
\centering
\includegraphics[width=0.9\columnwidth]{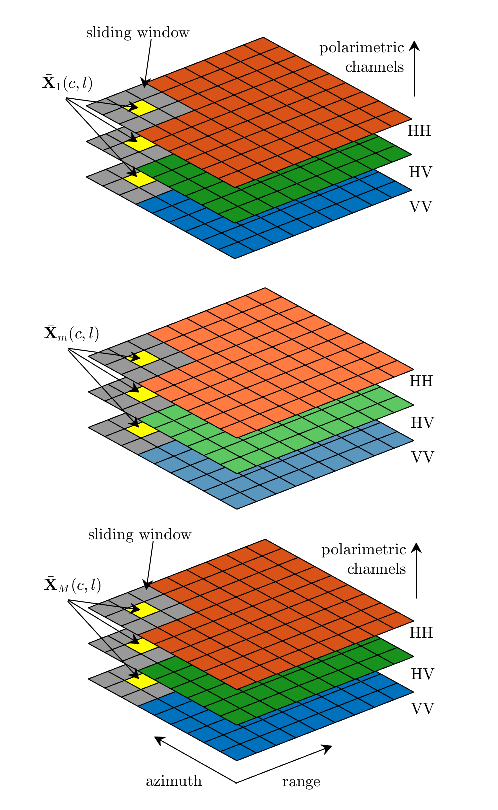}
\caption{A visual representation depicting the structure of the multipass \ac{PolSAR} datacube, comprising $M$ \ac{PolSAR} datacubes.}
\label{fig_DataCube}
\end{figure}

Now, in the presence of multiple observations (repeated passes) of the same scene, let us now consider the availability of $M$ datacubes, $\bY_m$, $m=1,\ldots,M$, each of them associated with the acquisition of a \ac{PolSAR} image representing the same scene. In the following, it is assumed that the $M$ \ac{PolSAR} images are already registered to each other. The formation process of the \ac{PolSAR} datacubes is hence described through a graphical representation in Figure \ref{fig_DataCube}.

Now, for each pixel under test, a rectangular neighborhood $\cal A$ of size $K = W_1 \times W_2$ is extracted, and the matrix whose columns are the vectors of the polarimetric returns from the pixels of $\bY_m$, $m = 1, \ldots, M$, which fall in the region $\cal A$ is denoted by $\bar{\bX}_m$, $m = 1, \ldots, M$. Let us also indicate with $\bX$ the multipass datacube obtained stacking together all single datacubes $\bar{\bX}_m$, $m=1,\ldots,M$, such that the information from a given pixel are collected along a specific column of $\bX$. Specifically, the pixels with the same coordinates are aligned to each other. Hence, the vectors contained in the sliding window centered on the pixel of interest are $\bx_1, \bx_2, \ldots, \bx_K$. In particular, each polarimetric vector $\bx_k$, $k=1,\ldots,K$ is obtained stacking the different vectors obtained in the $M$ passes. This implies that the size of each vector $\bx_k$, $k=1,\ldots, K$, is equal to $3 M$.

\section{Background on polarimetric covariance symmetries}\label{sec_background}

This section describes the polarimetric covariance matrix structures that arise when significant scattering symmetric properties \cite[pp. 69-72]{PottierBook}, \cite{Nghiem1992} dominate the return associated with the pixel under investigation. They are distinguished in absence of symmetry, reflection symmetry, rotation symmetry, and azimuth symmetry \cite{Nghiem1992}, whose structures are detailed in what follows.

The first considered structure examines the polarimetric covariance matrix in the presence of a reciprocal medium  \cite{PottierBook, Nghiem1992}, which is a $3 \times 3$ Hermitian matrix of the form 

\begin{equation}
\bC_{p} = \left[
\begin{array}{ccc}
c_{hhhh} & c_{hhhv} & c_{hhvv} \\
c_{hhhv}^* & c_{hvhv} & c_{hvvv} \\
c_{hhvv}^* & c_{hvvv}^* & c_{vvvv}
\end{array}
\right].
\end{equation}

This matrix depends on $\zeta_p = 9$ real-valued scalar quantities. In fact, its diagonal entries are the conventional (real-valued) backscattering coefficients, whereas the off-diagonal elements are complex.

The next structure is a reflection symmetry relative to a vertical plane. It is usually observed over horizontal natural environments and produces complete decorrelations between co-polarized and cross-polarized elements \cite{Nghiem1992}. Under reflection symmetry, the covariance matrix has a special form ruled by $\zeta_p = 5$ real scalar values, that is 

\begin{equation}
\bC_{p} = \left[
\begin{array}{ccc}
c_{hhhh} & 0 & c_{hhvv} \\
0 & c_{hvhv} & 0 \\
c_{hhvv}^* & 0 & c_{vvvv}
\end{array}
\right].
\end{equation}

Reflection symmetry can be observed, for instance, on water surfaces, in plowed fields, as well as in natural environments like forests, snow-covered areas, or sea ice. Additionally, this symmetry can be seen in structures like horizontal and vertical wires, or horizontal dihedrals \cite{Nghiem1992,PottierBook,pallotta2022covariance}.

Let us now consider the case of rotation symmetry \cite{Nghiem1992}. Under this assumption, the polarimetric covariance matrix assumes the following special structure

\begin{equation}
\bC_{p} = \left[
\begin{array}{ccc}
c_{hhhh} & c_{hhhv} & c_{hhvv}\\
-c_{hhhv} & c_{hvhv} & c_{hhhv}\\
c_{hhvv} & -c_{hhhv} & c_{hhhh} %
\end{array}
\right],
\end{equation}
with $c_{hhhv}$ purely imaginary, $c_{hhvv}$ purely real, and $c_{hvhv} = (c_{hhhh}-c_{hhvv})/2$. Under rotation symmetry, the covariance is hence described by $\zeta_p =3$ real scalar values. Rotation symmetry could be observed in forest canopy with helical structures in the vegetation (e.g., highly random vegetation) \cite{Nghiem1992, PottierBook, pallotta2022covariance, SHI2024209}.

The last considered symmetry, typically referred to as azimuth symmetry, is a combination of a rotation and a reflection symmetry \cite{Nghiem1992}. The polarimetric covariance matrix can be constructed from $\zeta_p =2$ real scalar values and has the form

\begin{equation}
\label{eq_azimuth}
\bC_{p} = \left[
\begin{array}{ccc}
c_{hhhh} & 0 & c_{hhvv} \\
0 & c_{hvhv} & 0 \\
c_{hhvv} & 0 & c_{hhhh}
\end{array}
\right],
\end{equation}
with $c_{hvhv} = (c_{hhhh}-c_{hhvv})/2$. An example of azimuth symmetry can be seen in vegetated areas at low frequencies and normal incidence, where the electromagnetic wave penetrates the foliage canopy and is scattered back by the horizontal branches or vertical tree trunks \cite{Nghiem1992,PottierBook,pallotta2022covariance}.

In what follows, ${\cal{P}}^S_{i}$, $i=1,\ldots,4$, is used to denote the sets of the above defined structured covariance matrices. 

\section{Problem formulation and proposed solution}\label{problem_formulation}

The polarimetric acquisitions of the same scene at different times may show total or partial decorrelation. This can happen due to changes in the object's geometry and its variability over time, as well as system decorrelation. Generally, the noise system is assumed negligible in the devised models, hence the available SAR images are acquired with high \ac{SNR}, and the decorrelation becomes a defining characteristic of the observed object only. Hence, in the following, multitemporal polarimetric information is properly exploited to effectively characterize the observed scene or objects in it contained.

Let us hence consider the availability of $K$ multitemporal polarimetric data $\bx_k$, $k=1,\ldots,K$, modelled as statistically \ac{i.i.d.} complex circular zero-mean Gaussian vectors with positive definite covariance matrix $\bC = \bC_t \otimes \bC_p$, namely

\begin{equation}{\label{eq_cronicker}}
\bx_k \sim \mathcal{CN}\left(0, \bC_t \otimes \bC_p\right),
\end{equation}
where $\bC_t$ is the $M \times M$ (unstructured) temporal covariance matrix related to multipass acquisitions, whereas $\bC_p$ is the $3 \times 3$ polarimetric covariance matrix that can exhibit one among the different structures described in Section \ref{sec_background}.\footnote{Notice that, the polarimetric matrix refers to the covariance (also known as coherency information) of polarimetric data gathered in a given acquisition (in time and space), modeling the scattering properties of the target, whereas, the temporal matrix accounts for the statistical dependency of the temporal data acquired in a given channel, and models the time-domain dynamics of the observed scene.}

Before proceeding further, let us discuss the reason behind the choice of this particular covariance structure. When dealing with high-dimensional covariances of multivariate phenomena, using Kronecker product decompositions allows estimating (lower-dimensional) covariances for each dimension \cite{Werner2008}. This Kronecker structure arises when the correlation function between two distinct groups of 2D (time and polarization) data can be factorized to separately represent intra-group and inter-group correlations \cite{cressie2011statistics}.

Let us now focus on the \ac{pdf} of the observable matrix $\bX$ \cite{Werner2008}

\begin{equation}\label{pdf_eq2}
f_{\bm{X}}(\bX|\bC) = \frac{1}{\pi^{N M K}\, {\left|\bC\right|}^K}\, \exp\left\{-\tr\left(\bC^{-1}\bX\bX^H\right)\right\},
\end{equation}
and denote by 

\begin{equation}
\hat\bS = \frac{1}{K} \bX \bX^H = \frac{1}{K} \sum_{k=1}^{K} \bx_k \bx_k^H,
\end{equation}
the \ac{SCM}. The corresponding negative log-likelihood function of $\bX$ can also be derived as

\begin{equation}\label{pdf_eq11}
l(\bC; \bX) = K \log |\bC| + K \tr \{\hat\bS  \bC^{-1} \} + NMK\log(\pi).
\end{equation}

Hence, assuming a specific structure for the polarimetric covariance, the \ac{ML} estimate of the covariance matrix $\bC$ can be obtained as the optimal solution to the following optimization problem under the constraints $\bC = \bC_t \otimes \bC_p$, $\bC_t \in {\cal{H}}_{M}^{++}$, and $\bC_p \in {\cal{P}}^S_{i} \cap {\cal{H}}_{N}^{++}$, $i=1,\ldots,4$, namely

\begin{equation}\label{pdf_eq12a}
\left\{
\begin{array}{ll}
\displaystyle{\min_{\bm{C}}} & l(\bC; \bX)\\
\text{s.t.} & \bC = \bC_t \otimes \bC_p\\
 & \bC_t \in {\cal{H}}_{M}^{++}\\
 & \bC_p \in {\cal{P}}^S_{i} \cap {\cal{H}}_{N}^{++},\; i=1,\ldots, 4
\end{array}
\right.
\end{equation}
or equivalently 

\begin{equation}\label{pdf_eq12}
\left\{
\begin{array}{ll}
\displaystyle{\min_{\bm{C}}} & \log |\bC| +  \tr \{\hat\bS  \bC^{-1} \}\\
\text{s.t.} & \bC = \bC_t \otimes \bC_p\\
 & \bC_t \in {\cal{H}}_{M}^{++}\\
 & \bC_p \in {\cal{P}}^S_{i} \cap {\cal{H}}_{N}^{++},\; i=1,\ldots, 4
\end{array}
\right.
\end{equation}

When a structure is imposed on the covariance matrices in addition to being Hermitian (as for instance the Kronecker structure), the optimization problem in \eqref{pdf_eq12} cannot in general be solved in closed form. As a consequence, its solution (leading to an estimate of the two covariances $\bC_t$ and $\bC_p$) can be approached, with some guarantee of optimality, through an iterative alternating minimization algorithm (a.k.a. flip-flop or ping-pong) \cite{Werner2008}. As a matter of fact, it involves alternating minimization with respect to $\bC_t$ and $\bC_p$, while keeping the most recent estimate of $\bC_t$ fixed when minimizing over $\bC_p$ and viceversa \cite{Werner2008}.

To this end, let us recall some useful properties related to the Kronecker product of matrices. Knowing the relationship between the determinant of the Kronecker product \( \mathbf{C}_t \otimes \mathbf{C}_p \) and the determinants of the matrices \( \mathbf{C}_t \) and \( \mathbf{C}_p \) is given by

\[
\left|\mathbf{C}_t \otimes \mathbf{C}_p\right| = \left|\mathbf{C}_t\right|^M \cdot \left|\mathbf{C}_p\right|^N
\]

Exploiting the above result, the objective function in \eqref{pdf_eq12} can be recast as

\begin{equation}\label{pdf_eq}
\begin{split}
l(\bC_t,\bC_p; \bX) &= M \log |\bC_t| + N \log |\bC_p|\\ 
&\quad + \tr \left\{\hat\bS \left(\bC_t^{-1}\otimes \bC_p^{-1}\right) \right\}.
\end{split}
\end{equation}

\noindent
\textbf{Optimization over $\bC_p$}. For a fixed $\bC_t$, the \ac{ML} estimate of $\bC_p$ can be obtained by minimizing \eqref{pdf_eq} while enforcing one of the special structures described in Section \ref{sec_background}. More specifically, the minimizer of \eqref{pdf_eq} with respect to $\bC_p \in {\cal{P}}^S_{i} \cap {\cal{H}}_{N}^{++}$ can be found as the optimal solution to the following constrained minimization problem:

\begin{equation}\label{pdf_eq_single}
\left\{
\begin{array}{ll}
\displaystyle{\min_{\bm{C}_p}} & \log |\bC_p| + \tr \{\hat{\bC}_p \bC_p^{-1} \}\\
\text{s.t.} & \bC_p \in {\cal{P}}^S_{i} \cap {\cal{H}}_{N}^{++},\; i=1,\ldots, 4
\end{array}
\right.
\end{equation}
where 
\begin{equation}\label{OptimalCpForfixedCt_eq}
   \hat{\bC}_p(\bC_t) = \frac{1}{M} \sum_{k=1}^{M}\sum_{l=1}^{M} \hat{\bS}^{kl} \bC_t^{-1}(l,k),
\end{equation}
with $\hat{\bS}^{kl}$ the $(k,l)$-th block of size $N \times N$ in the matrix $\hat\bS$, and $\bC_t^{-1}(l,k)$ the $(k,l)$-th element of the matrix $\bC_t^{-1}$. Note that Problem \eqref{pdf_eq_single} is formally analogous to the problem presented in \cite[eq. 11]{pallotta2016detecting}, with $\hat{\bC}_p$ replacing the SCM. Consequently, the principal technical result of \cite{pallotta2016detecting} can be applied to derive the \ac{ML} estimation of the structured polarimetric covariance matrix $\bC_p$. Thus, the optimal solution $\breve{\bC}_p$ is provided below for each specified structural constraint, using the matrix statistic $\hat{\bC}_p$ as starting point.

\begin{enumerate}
\item No symmetry:
\begin{equation}
    \breve{\bC_p} = \hat{\bC}_p.
    \label{eq_Cp_NoSymmetry}
\end{equation}

\item Reflection symmetry:
\begin{equation}
    \breve{\bC_p} = \bU^{H} \left[\begin{array}{cc}
\tilde{\bS}_{1,1} & \bzero\\
\bzero & \tilde{S}_{3,3}
\end{array} \right] \bU
\label{eq_Reflection_symmetry}
\end{equation}
where $\tilde{\bS} = \bU\hat{\bC}_p\bU^{H} = \left[\begin{array}{cc}
\tilde{\bS}_{1,1} & \tilde{\bS}_{1,3}\\
\tilde{\bS}_{3,1} & \tilde{S}_{3,3}
\end{array} \right]$, and 
$$
\bU = \left[
\begin{array}{ccc}
1 & 0 & 0 \\
0 & 0 & 1 \\
0 & 1 & 0
\end{array}
\right].
$$
\item Rotation symmetry:
\begin{equation}\label{eq_Rotation_symmetry}
\begin{split}
\breve{\bC_p} = \bT^{H} &\bE^{-1} \bV^{H} \left[
\begin{array}{cc}
\tilde{S}_{1,1} & \bzero\\
\bzero & \frac{1}{2}\left(\tilde{\bS}_{2,2} +\bJ \tilde{\bS}_{2,2} \bJ\right)
\end{array}
\right] \\
&\bV \bE^{-1} \bT
\end{split}
\end{equation}

where $\tilde{\bS} = \bV\bE\bT\hat{\bC}_p\bT^{H}\bE\bV^{H} = \left[
\begin{array}{cc}
\tilde{S}_{1,1} & \tilde{\bS}_{1,2}\\
\tilde{\bS}_{2,1} & \tilde{\bS}_{2,2}
\end{array}
\right]$, and
$$
\bE = \left[
\begin{array}{ccc}
1 & 0 & 0 \\
0 & 1/\sqrt{2} & 0 \\
0 & 0 & 1
\end{array}
\right],\;
\bT = \frac{1}{\sqrt{2}}\left[
\begin{array}{ccc}
1 & 0 & 1 \\
1 & 0 & -1 \\
0 & \sqrt{2} & 0
\end{array}
\right],
$$
$$
\bV = \left[
\begin{array}{ccc}
1 & 0 & 0 \\
0 & 0 & j \\
0 & 1 & 0
\end{array}
\right].
$$
\item Azimuth symmetry:
\begin{equation}\label{eq_azimuth_symmetry}
\begin{split}
\breve{\bC_p} = \bT^{H} &\bE^{-1} \diag\left(\tilde{S}_{1,1}, \frac{\tilde{S}_{2,2} + \tilde{S}_{3,3}}{2}, \frac{\tilde{S}_{2,2} + \tilde{S}_{3,3}}{2}\right) \\
&\bE^{-1} \bT
\end{split}
\end{equation}

where $\tilde{\bS} = \bE\bT\hat{\bC}_p\bT^{H}\bE$ and $\tilde{S}_{1,1}$, $\tilde{S}_{2,2}$, $\tilde{S}_{3,3}$ are its diagonal entries.
\end{enumerate}

\noindent
\textbf{Optimization over $\bC_t$}. As in the previous case, for a fixed $\bC_p$, the minimizer of \eqref{pdf_eq} with respect to $\bC_t \in {\cal{H}}_{M}^{++}$ can be found as the optimal solution to the following minimization problem:

\begin{equation}\label{pdf_eq_single2}
\left\{
\begin{array}{ll}
\displaystyle{\min_{\bm{C}_t}} & \log |\bC_t| + \tr \{\hat{\bC}_t \bC_t^{-1} \}\\
\text{s.t.} & \bC_t \in {\cal{H}}_{M}^{++}
\end{array}
\right.
\end{equation}
whose optimal solution is given by

\begin{equation}
   \hat{\bC}_t(\bC_p)= \frac{1}{N} \sum_{k=1}^{N}\sum_{l=1}^{N} \bar\bS^{kl} \bC_p^{-1}(l,k)
   \label{OptimalCtForfixedCp_eq}
\end{equation}
where
\begin{equation}
    \bar\bS= \bK^T_{M,N} \hat{\bS} \bK_{M,N}
\end{equation}
and $\bar\bS^{kl} $ is the $(k,l)$-th block of $M \times M$ in the matrix $\bar\bS$. The permutation matrix $\bK_{M,N}$ is defined such that

\begin{equation}
    \bK_{M,N} \text{vec}\{\bA\} = \text{vec}\{\bA^T\},
\end{equation}
for any \( M \times N \) matrix $\bA$, and its explicit form is given by
\begin{equation}
    \bK_{M,N} = \sum_{i=1}^{M} \sum_{j=1}^{N} (E_{ij} \otimes E_{ji}),
    \end{equation}
where $E_{ij}$ is an $M \times N$ matrix with 1 in the \((i,j)\)-th position and 0 elsewhere. For instance, for $N=3$ and $M=2$, it is
\begin{equation}
  \bK_{2,3} =  \begin{pmatrix}
1 & 0 & 0 & 0 & 0 & 0 \\
0 & 0 & 1 & 0 & 0 & 0 \\
0 & 0 & 0 & 0 & 1 & 0 \\
0 & 1 & 0 & 0 & 0 & 0 \\
0 & 0 & 0 & 1 & 0 & 0 \\
0 & 0 & 0 & 0 & 0 & 1
\end{pmatrix}
\end{equation}

Note that, since $\hat{\bS}$ is a psd matrix, also, $\hat{\bC}_p$ in \eqref{OptimalCpForfixedCt_eq} and $\hat{\bC}_t$ in \eqref{OptimalCtForfixedCp_eq} are psd matrices \cite{Werner2008}.

\vspace{0.25cm}
To solve the problem in \eqref{pdf_eq12}, the flip-flop which involves alternating minimization with respect to $\bC_t$ and $\bC_p$ (keeping one of them fixed at a time) is hence applied. Its main steps are reported in Algorithm \ref{flipflop_alg2}.

\begin{algorithm}
\caption{Flip-flop algorithm with polarimetric symmetry structure}
\label{flipflop_alg2}
\begin{algorithmic}[1]
    \REQUIRE $K$, $\bX$ (constructed as described in Section \ref{ProbFormulation}).
    \ENSURE $\breve{\bC}_p$, $\hat{\bC}_t$ for each  symmetry.
    \STATE Initialize $\hat{\bC}_t$ = $\bC_t^{init}$ by an arbitrary psd matrix.
    \STATE Calculate $\hat{\bC}_p$ using \eqref{OptimalCpForfixedCt_eq} with the previous $\hat{\bC}_t$.
    \STATE Force the symmetry structure, by calculating the     \ac{ML} estimate $\breve{\bC_p}$ as in \eqref{eq_Cp_NoSymmetry}, \eqref{eq_Reflection_symmetry}, \eqref{eq_Rotation_symmetry} or \eqref{eq_azimuth_symmetry} using $\hat{\bC}_p$ as starting point.
    \STATE Set $\hat{\bC}_p = \breve{\bC_p}$. 
    \STATE Calculate the new value of $\hat{\bC}_t$ using \eqref{OptimalCtForfixedCp_eq}, with the updated $\hat{\bC}_p$ in the previous step.
    \STATE Iterate steps 2-5 until convergence.
\end{algorithmic}
\end{algorithm}

\begin{remark} 
{\normalfont It is worth noting that, the proposed flip-flop estimation procedure (assuming the existence of the \ac{ML} estimate) exhibits the following feature properties \cite{razaviyayn2013unified, aubry2018sequential}:
\begin{itemize}
    \item the sequence of produced negative log-likelihood values is bounded below;
    \item the sequence of produced negative log-likelihood values is monotonically decreasing and converges to a finite value. Moreover, for any cluster point, the corresponding negative log-likelihood value attains this limit value;
    \item any cluster point of the produced sequence of estimates is a stationary solution to Problem \eqref{pdf_eq} and the corresponding negative log-likelihood value attains this aforementioned limit value.
\end{itemize}}
\end{remark}

\section{Classification of dominant symmetry by model order selection}\label{sec_MOS}

The framework outlined in this section identifies the predominant special structure of the polarimetric covariance matrix characterizing the scattering mechanism of a given pixel. This allows the problem to be treated as a data classification task based on the scattering properties of the target pixel and its neighboring pixels. As a result, the classification problem is formulated as a multiple hypothesis test that includes both nested and non-nested hypotheses, that is:

\begin{equation}
\left\{
\begin{array}{ll}
H_1: & \text{no symmetry;}\\
H_2: & \text{reflection symmetry;}\\
H_3: & \text{rotation symmetry;}\\
H_4: & \text{azimuth symmetry.}
\end{array}
\right.
\end{equation}

The decision is taken resorting to \ac{MOS} criteria \cite{Selen2004, stoica2004model}, selecting the hypothesis $H_{i^{\star}}$ according to

\begin{equation}\label{eq_selection_order}
 i^{\star} = \text{arg}\,\displaystyle{\min_{i=1,\ldots,4}}\; \left[ 2 l\left(\hat{\bC}_{(i)}; \bX\right) + \zeta \eta(K)\right],
\end{equation}
where $\hat{\bC}_{(i)}$ is the \ac{ML} estimate of $\bC$ under the $H_i$ hypothesis comprising $\zeta = \zeta_t + \zeta_p$ parameters, with $\zeta_t=M^2$, whereas $\zeta_p$ are as detailed in Section \ref{sec_background}. The quantity $\zeta\,\eta(K)$ represents the penalty coefficient used to prevent overfitting \cite{stoica2004model}. Indeed, as the negative log-likelihood in \eqref{eq_selection_order} decreases with increasing $\zeta$ (for nested models), the second term correspondingly increases. Moreover, different selection strategies diversify due to the definition of the penalty, namely:
\begin{itemize}
\item AIC (Akaike information criterion): $\eta(K) = 2$;
\item GIC (generalized information criterion): $\eta(K) = \delta+1$, with $\delta$ an integer number greater than or equal to $2$;
\item BIC (Bayesian information criterion): $\eta(K) = \log(K)$;
\item HQC (Hannan-Quinn information criterion) \cite{hannan1979determination}: $\eta(K) = 2\log(\log(K))$.
\end{itemize}

Finally, taking into account the statistical hypotheses regarding the data, i.e., by substituting the expression of $l\left(\hat{\bC}_{(i)}; \bX\right)$ in \eqref{eq_selection_order}, the synthesized \ac{MOS} rules become

\begin{equation}
i^{\star} = \text{arg}\,\displaystyle{\min_{i=1,\ldots,4}}\,\left[
2K\log \left|\hat{\bC}_{(i)}\right| + 2K\, \tr\left({\hat{\bC}^{-1}}_{(i)}\hat{\bS}\right) + \zeta\, \eta(K)\right]
\label{eq_MOS} 
\end{equation}

A summary of the generic classification scheme is illustrated in Algorithm \ref{mos_alg}.

\begin{algorithm}
\caption{MOS algorithm for symmetry classification}
\label{mos_alg}
\begin{algorithmic}[1]
    \REQUIRE $\hat{\bC}_t$, $\hat{\bC}_p$ for each specific symmetry (constructed as described in Section \ref{problem_formulation}).
    \ENSURE $H_{i^{\star}}$, i.e., the selected dominant symmetry.
    \STATE Calculate the decision statistic with \eqref{eq_MOS} using the founded $\hat{\bC}_t$ and $\breve{\bC_p}$ for each symmetry, and a specific \ac{MOS} criterion, viz., AIC, GIC, BIC or HQC.
    \STATE Select the hypothesis (resp. symmetry) which corresponds to the minimum decision statistic.
\end{algorithmic}
\end{algorithm}

\section{Performance analysis and discussions}\label{secPerformanceAnalysisAndDiscussions}

This section is intended to evaluate the effectiveness of the proposed polarimetric covariance symmetry classification framework for multitemporal \ac{PolSAR} images. To this end, both quantitative analyses on simulated data and assessments on measured \ac{PolSAR} data are performed.

\subsection{Quantitative analysis on simulated data}\label{sub_analysis_SimulatedData}

This subsection evaluates the performance of the proposed covariance symmetry classification scheme for multitemporal \ac{PolSAR} images. The first analysis is conducted to quantify the accuracy of the proposed flip-flip in estimating the polarimetric covariance matrix. Hence, the normalized \ac{RMSE} is used as performance metric, which is defined as 

\begin{equation}\label{RMSE}
\text{NRMSE} = \sqrt{\E\left[\frac{\left\|\bC_p - \breve{\bC}_p \right\|^2}{\left\|\bC_p \right\|^2}\right]}.
\end{equation}

Noteworthy, due to the lack of a closed-form expression for the normalized RMSE in \eqref{RMSE}, the statistical expectation is approximated resorting to the classic Monte Carlo approach, that is 

\begin{equation}\label{RMSE2}
\text{NRMSE} = \sqrt{\frac{1}{M_c}\sum_{m=1}^{M_c}\frac{\left\|\bC_p - \breve{\bC}_p^{(m)} \right\|^2}{\left\|\bC \right\|^2}},
\end{equation}
where $\breve{\bC}_p^{(m)}$ is the estimate of $\bC_p$ obtained at the $m$-th Monte Carlo run and $M_c$ is the number of trials. Moreover, as a term of comparison, the structured \ac{ML} covariance estimator assuming independent and identically distributed polarimetric data in different temporal looks, i.e., supposing the absence of temporal correlation, is considered. Precisely, the structured \ac{ML} covariance estimate is obtained applying the classification procedure of \cite{pallotta2016detecting}, where the \ac{SCM} is replaced by the average of the two \ac{SCM}s obtained from two distinct data subsets, each of size $K$. Hence, it is denoted as temporal uncorrelated structured \ac{ML} (TUSML) for short in the next analyses.

Differently, the second analysis concentrates on assessing the effectiveness of the \ac{MOS} rules devised in Section \ref{sec_MOS} in terms of its capabilities to correctly detect (or classify) the true symmetry. To this end, the accuracy (which also represents the probability of correctly selecting a specific structure) as well as the confusion matrix are introduced as a performance metric. Due to the lack of a closed-form expression for the accuracy, it is computed resorting to the classic Monte Carlo approach as the ratio between the number of times a specific symmetry is selected over the total number of trials, $M_c$, set equal to $10^4$.

The considered simulation scenario consists of $K$ zero-mean complex circular Gaussian vectors of size $3M$, which are colored with a specific nominal covariance. Specifically, the nominal covariance matrix for each of the considered four scenarios (no symmetry, reflection, rotation, and azimuth symmetry) is $\bC_i = \bC_t \otimes \bC_{p_i}$, $i=1,\ldots, 4$. Regarding the temporal covariance matrix, two different situations are studied, i.e., $\bC_t = \bI$ (uncorrelated data), and $\bC_t(n,m) = \rho^{|n-m|}$, $\{n,m\}= 1,\ldots, M$, $|\rho|<1$ (correlated data with exponentially shaped covariance structure). Differently, the nominal polarimetric covariances for the four scenarios are defined, respectively, as \cite{pallotta2016detecting}

\begin{equation}\label{eq_Cp1}
\bC_{p_1} = \left[
\begin{array}{ccc}
1 & 0.2 + 0.3j & 0.5 - 0.3j\\
0.2 - 0.3j & 0.25 & -0.2 - 0.2j\\
0.5 + 0.3j & -0.2 + 0.2j & 0.8
\end{array}
\right],
\end{equation}
\begin{equation}\label{eq_Cp2}
\bC_{p_2} = \left[
\begin{array}{ccc}
1 & 0 & 0.5 - 0.3j\\
0 & 0.25 & 0\\
0.5 + 0.3j & 0 & 0.4
\end{array}
\right],
\end{equation}
\begin{equation}\label{eq_Cp3}
\bC_{p_3} = \left[
\begin{array}{ccc}
1 & 0.3j & 0.2\\
- 0.3j & 0.4 & 0.3j\\
0.2 & - 0.3j & 1
\end{array}
\right],
\end{equation}
\begin{equation}\label{eq_Cp4}
\bC_{p_4} = \left[
\begin{array}{ccc}
1 & 0 & 0.5\\
0 & 0.25 & 0\\
0.5 & 0 & 1
\end{array}
\right].
\end{equation}

\begin{figure}[ht!]
    \centering
    \includegraphics[width=0.6\columnwidth]{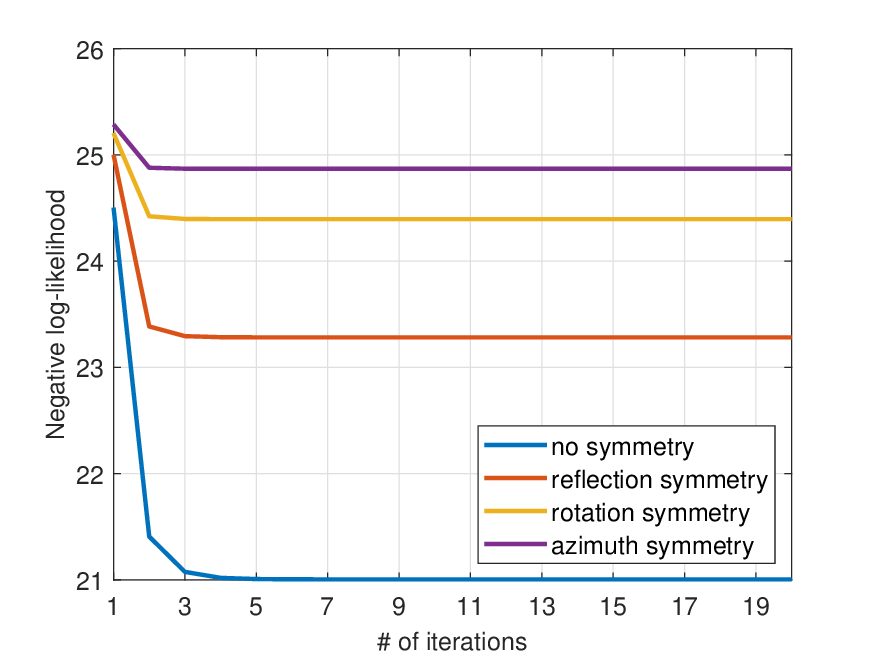}
    \caption{Negative log-likelihood (evaluated over $M_c=10^4$ Monte Carlo trials) versus the number of iterations in the flip-flop algorithm for $M=2$ simulated data with exponentially shaped $\bC_t$ with $\rho=0.9$, and $K=6$.}
    \label{convergence}
\end{figure}

All the analyses are conducted with the flip-flop algorithm initialed with the identity matrix for $\hat{\bC}_t$ and executed with $5$ iterations. This number is chosen since the negative log-likelihood in \eqref{pdf_eq} does not undergo significant variations beyond this amount of steps. As an example, the readers may refer to Figure \ref{convergence} for the above simulation scenario with $M=2$ and $K=6$.

\begin{figure}[ht!]
    \centering
    \subfigure[no symmetry]{
    \includegraphics[width=0.46\columnwidth]{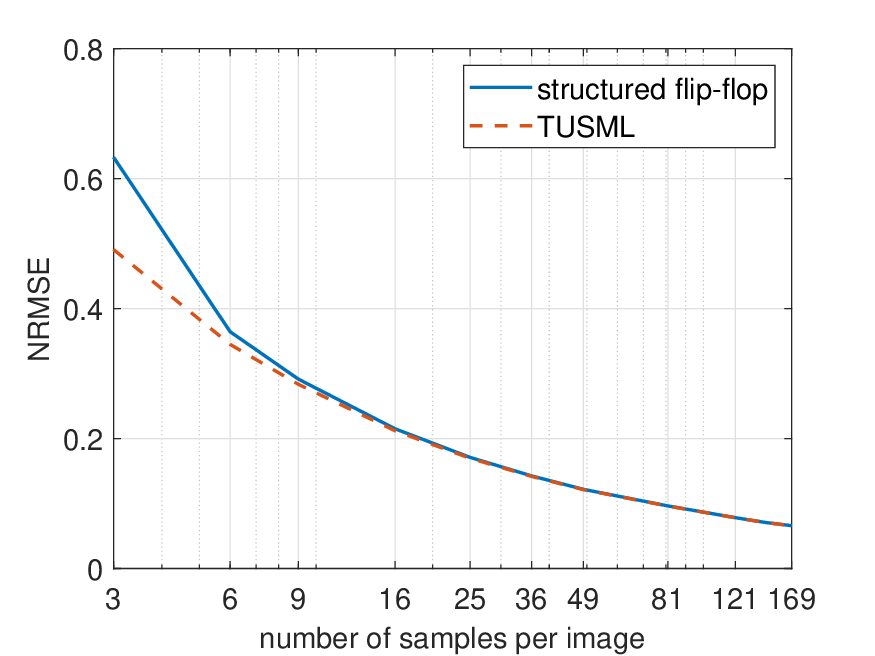}}
    \subfigure[reflection symmetry]{
    \includegraphics[width=0.46\columnwidth]{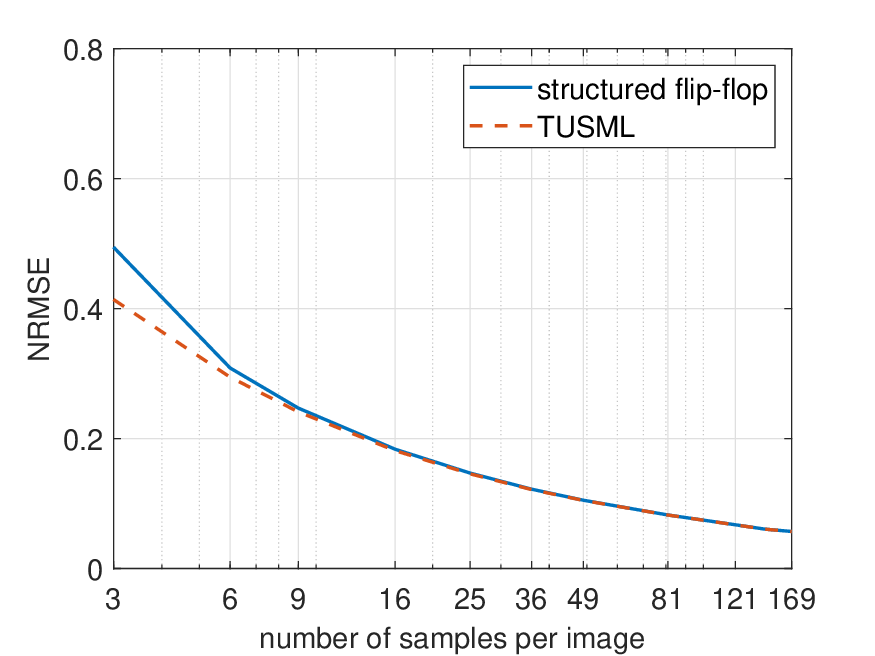}}\\
    \subfigure[rotation symmetry]{
    \includegraphics[width=0.46\columnwidth]{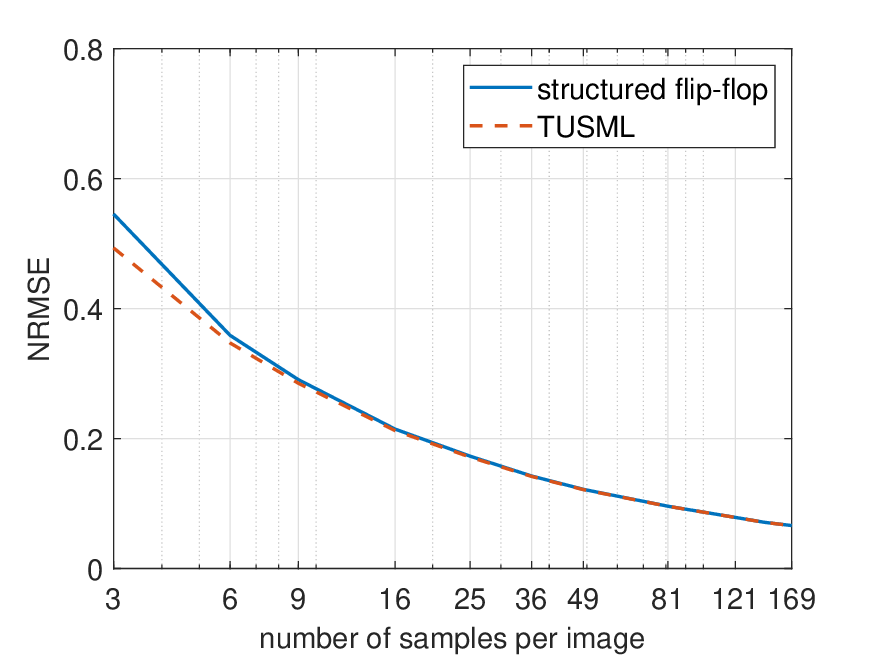}} 
    \subfigure[azimuth symmetry]{
    \includegraphics[width=0.46\columnwidth]{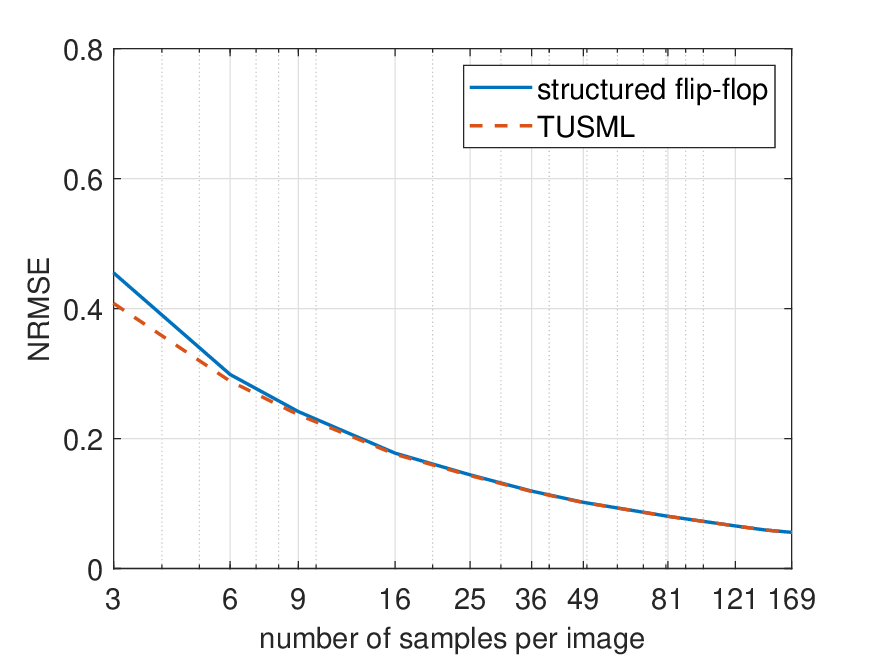}}
    \caption{Normalized RMSE (evaluated over $M_c=10^4$ Monte Carlo trials) of the polarimetric covariance estimate versus the number of samples per subset. The curves refer to the proposed flip-flop and the TUSML. Data are simulated according to the described Gaussian model setting $M=2$ and $\bC_t=\bI$. Subplots refer to the nominal $\bC_p$ having a) no symmetry, b) reflection symmetry, c) rotation symmetry, and d) azimuth symmetry.}
    \label{RMSEvsK_2imagesCtId}
\end{figure}

\begin{figure}[ht!]
    \centering
    \subfigure[no symmetry]{
    \includegraphics[width=0.46\columnwidth]{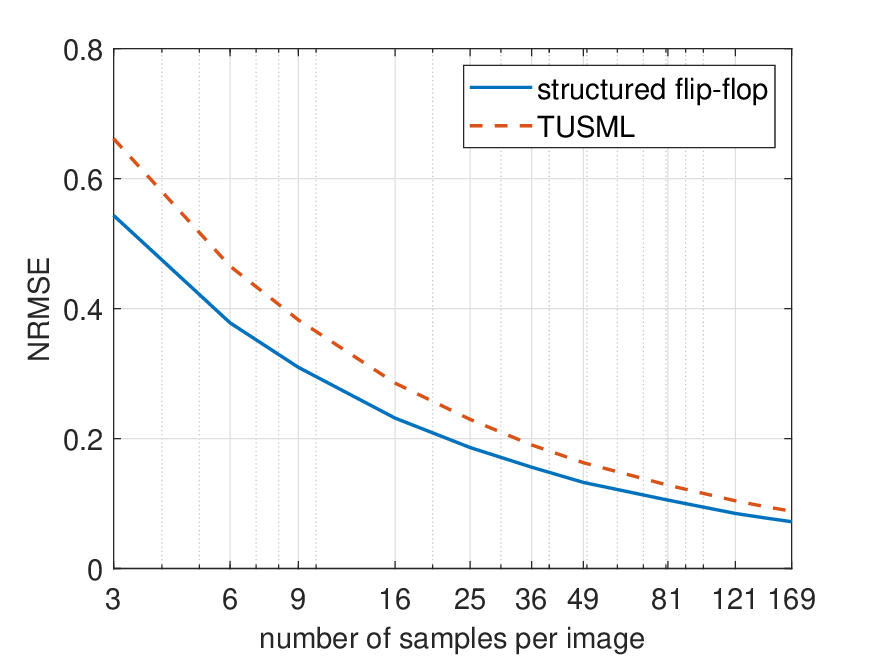}}
    \subfigure[reflection symmetry]{
    \includegraphics[width=0.46\columnwidth]{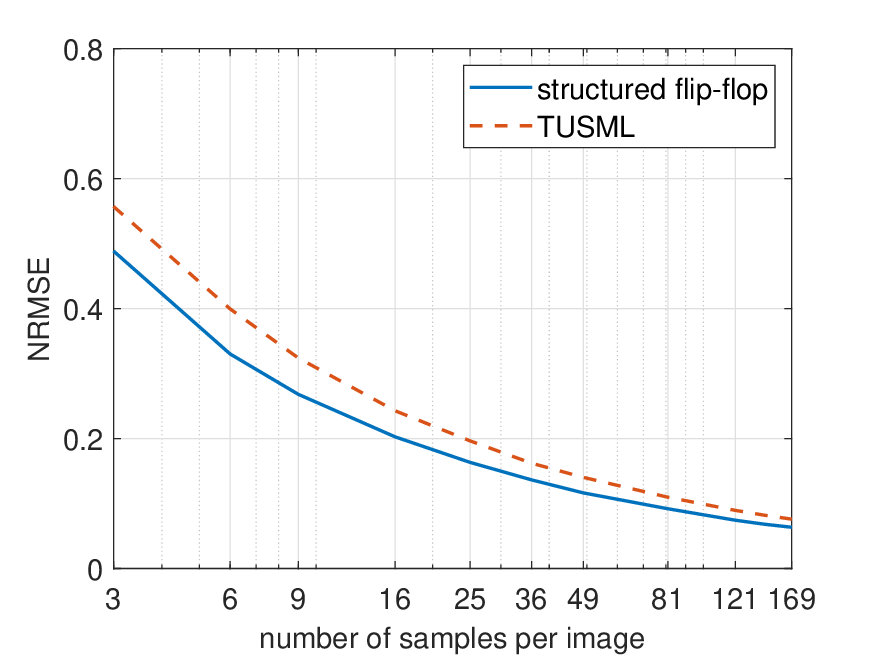}}\\
    \subfigure[rotation symmetry]{
    \includegraphics[width=0.46\columnwidth]{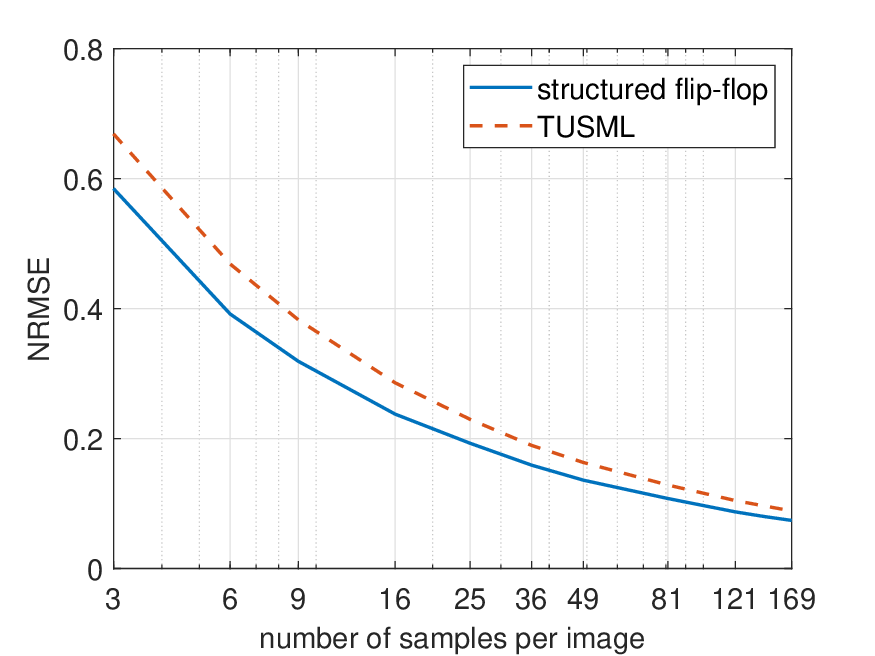}} 
    \subfigure[azimuth symmetry]{
    \includegraphics[width=0.46\columnwidth]{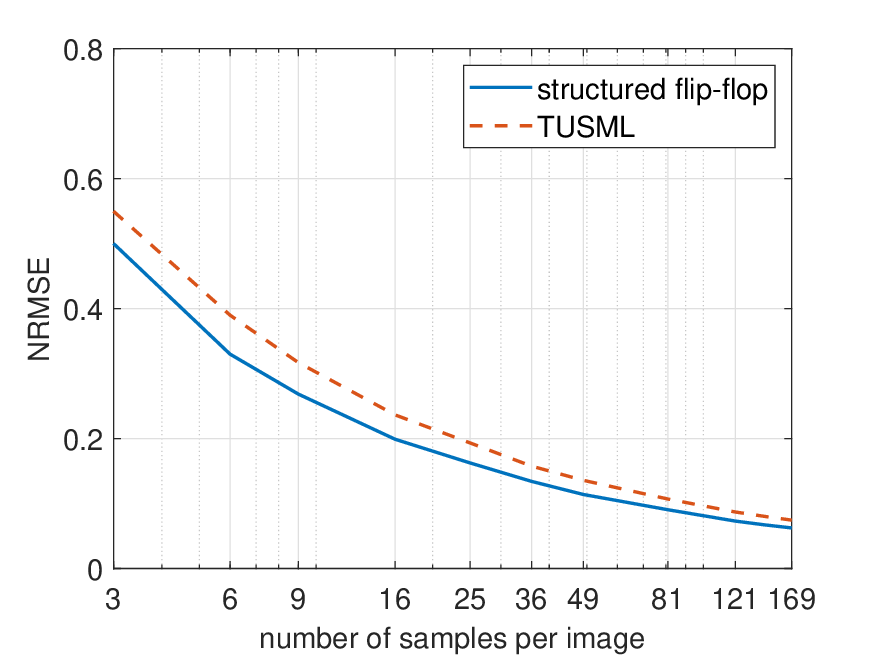}}
    \caption{Normalized RMSE (evaluated over $M_c=10^4$ Monte Carlo trials) of the polarimetric covariance estimate versus the number of samples per subset. The curves refer to the proposed flip-flop and the TUSML. Data are simulated according to the described Gaussian model setting $M=2$ and exponentially shaped $\bC_t$ with $\rho = 0.9$. Subplots refer to the nominal $\bC_p$ having a) no symmetry, b) reflection symmetry, c) rotation symmetry, and d) azimuth symmetry.}
    \label{RMSEvsK_2imagesCtrho0.9}
\end{figure}

\noindent
\textbf{Normalized RMSE analysis.} The normalized RMSE is shown versus the number of samples in Figure \ref{RMSEvsK_2imagesCtId} for the proposed flip-flop and the TUSML. The simulated scenario assumes $\bC_t = \bI$, namely temporal uncorrelation, and $M = 2$, i.e., two temporal acquisitions, where subplots refer to different structures for $\bC_p$ (both in terms of data generation and adopted structured \ac{ML} estimate). Similarly, Figure \ref{RMSEvsK_2imagesCtrho0.9} shows the normalized RMSE versus the number of samples for the same simulation setting, except for the introduction of temporal correlation, i.e., $\bC_t$ is exponentially shaped with $\rho = 0.9$. Interestingly, for the scenario of Figure \ref{RMSEvsK_2imagesCtId} assuming temporal independence, the TUSML slightly overcomes the proposed flip-flop in the presence of a reduced number of spatial samples, for all the considered structures. While, in all the other circumstances, namely in the presence of correlation between the data, the proposed flip-flop demonstrates its superiority. The observed behaviors can be explained by noting that, in the first case, the algorithm requires more data to achieve the same performance due to the high number of unknowns. In contrast, in the second case, while the TUSML experiences a model mismatch, the flip-flop method does not. Moreover, as expected, both the estimators under all the considered scenarios improve their performance as the number of looks increases. 

\begin{figure}[ht!]
    \centering
    \subfigure[AIC]{
    \includegraphics[width=0.46\columnwidth]{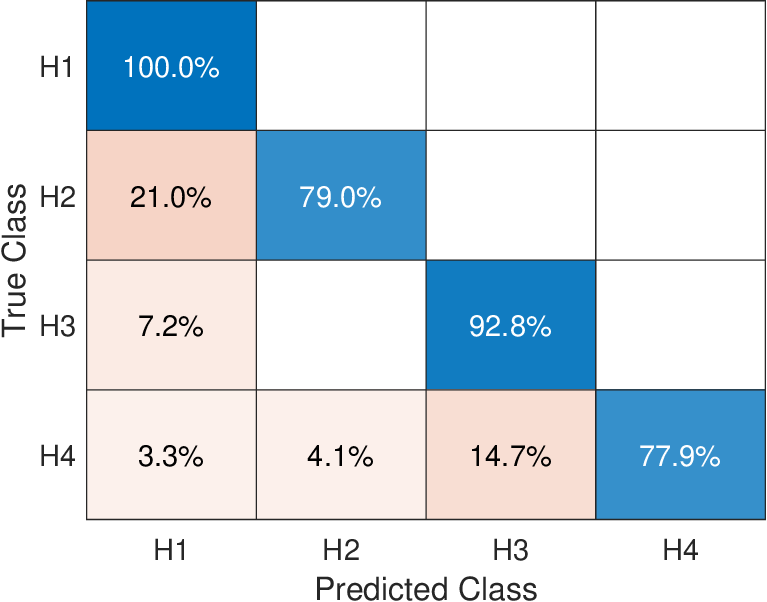}}
    \subfigure[BIC]{
    \includegraphics[width=0.46\columnwidth]{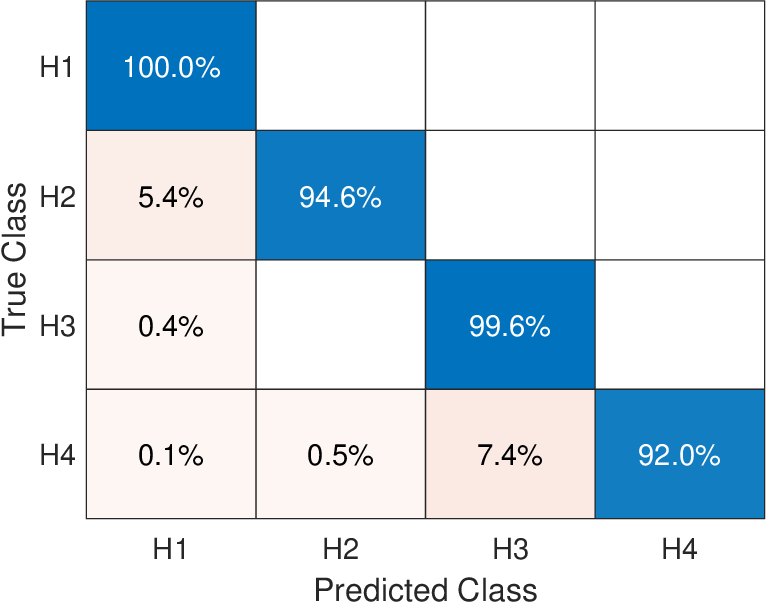}}\
    \subfigure[GIC with $\delta=2$]{
    \includegraphics[width=0.46\columnwidth]{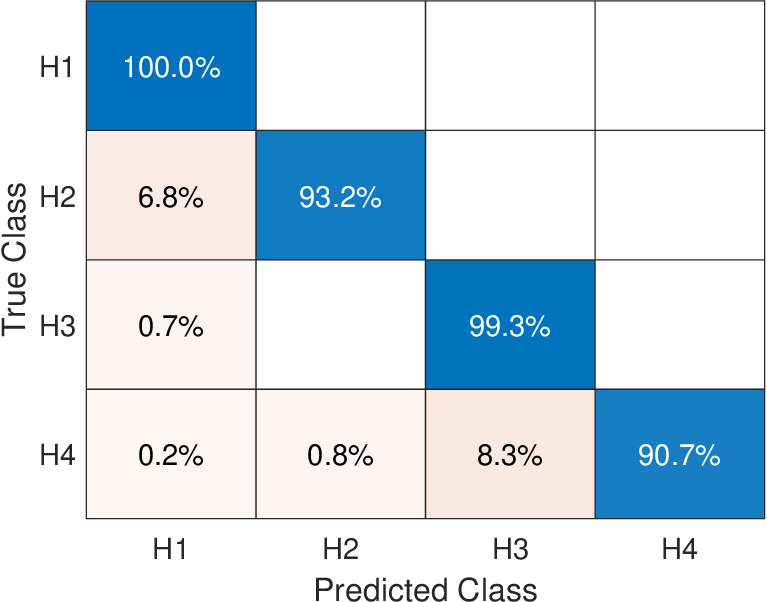}}
    \subfigure[HQC]{
    \includegraphics[width=0.46\columnwidth]{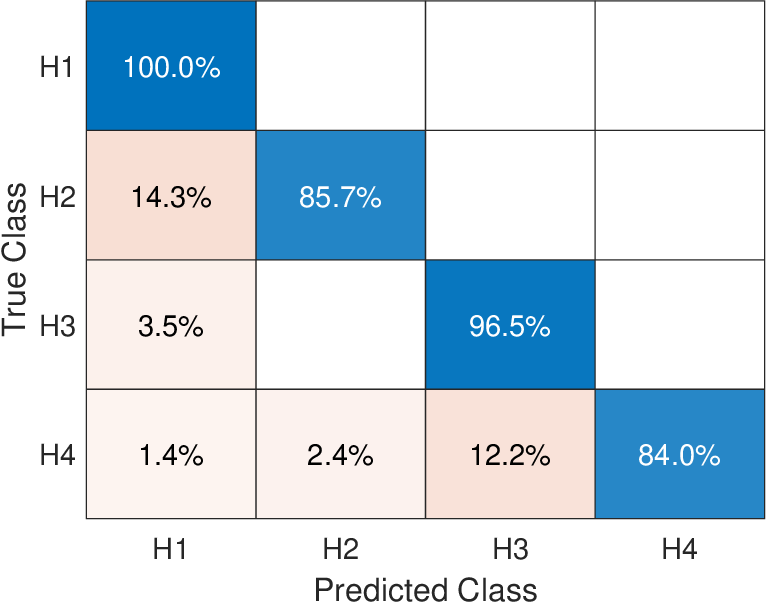}}
    \caption{Confusion matrix for the considered \ac{MOS} rules (for the classes of no symmetry, reflection symmetry, rotation symmetry, and azimuth symmetry), using  $K=25$ looks with $M=2$ and $\bC_t=\bI$. Subplots refer to the \ac{MOS} rules a) AIC, b) BIC, c) GIC, and d) HQC.}
    \label{Simu_2imagesK25}
\end{figure}

\vspace{0.25cm}
\noindent
\textbf{Classification capabilities analysis.} To further shed light on the capabilities of the proposed framework of correctly selecting the proper polarimetric covariance structure, an analysis in terms of accuracy is herein reported and discussed. In Figure \ref{Simu_2imagesK25}, the confusion matrix (expressed in percentage) is shown for data generated according to the above described Gaussian model for each one of the polarimetric structures in \eqref{eq_Cp1}-\eqref{eq_Cp4}, assuming $K=25$ looks, and $M=2$ temporal instances. Hence, each subplot in the figure refers to each of the \ac{MOS} rules described in Section \ref{sec_MOS}, viz., AIC, BIC, GIC (with $\delta=2$), and HQC. Moreover, the four classes are representative of the nominal polarimetric structure. Therefore, for each hypothesis (i.e., each nominal symmetry), the probability of selecting each possible structure is described by each row in the confusion matrix.

From the inspection of these results, it is evident that all the considered \ac{MOS} rules are in general capable of identifying the nominal structure in the polarimetric covariance. It also turns out that azimuth symmetry is the most difficult to identify, as it can be seen as a special case of all the other structures, as shown by the lowest accuracy values, for each considered \ac{MOS}. Nevertheless, both the BIC and GIC demonstrate to be the most effective classification rules, outperforming both the AIC and HQC counterparts. Moreover, they ensure an accuracy greater than the $90\%$ also for the challenging azimuth symmetry case. Interestingly, the BIC allows reaching the best performance without the necessity of setting any tuning parameter. 

\begin{figure}[ht!]
    \centering
    \subfigure[AIC]{
    \includegraphics[width=0.46\columnwidth]{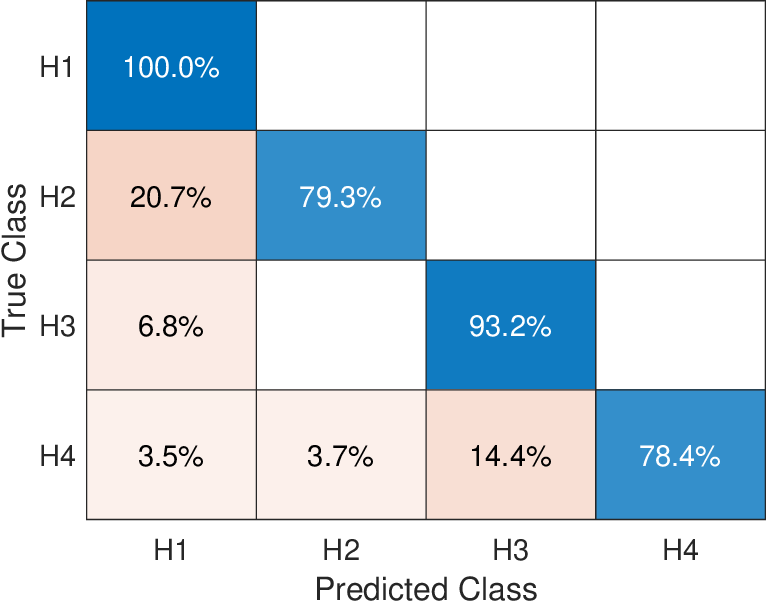}}
    \subfigure[BIC]{
    \includegraphics[width=0.46\columnwidth]{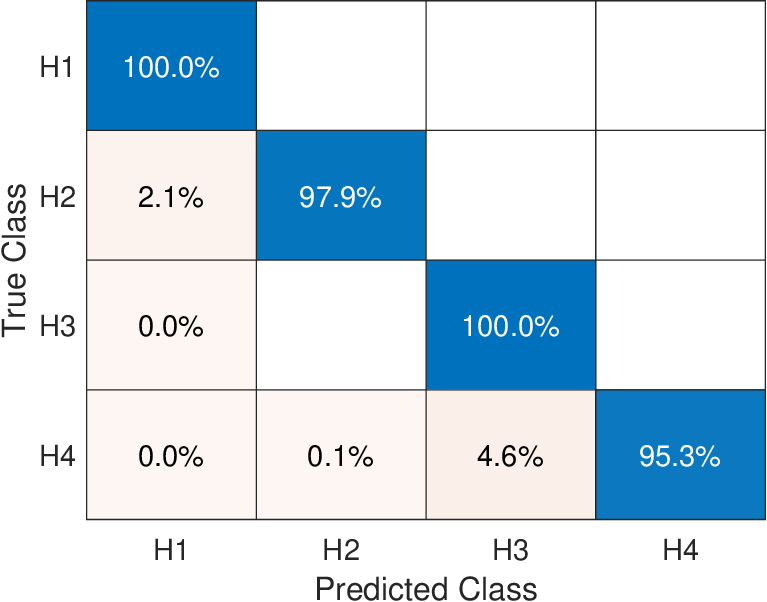}}\\
    \subfigure[GIC with $\delta=2$]{
    \includegraphics[width=0.46\columnwidth]{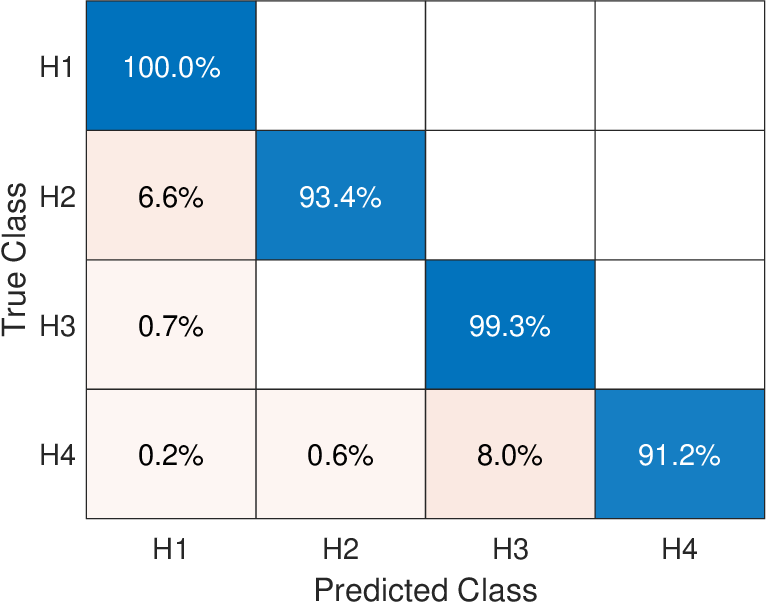}}
    \subfigure[HQC]{
    \includegraphics[width=0.46\columnwidth]{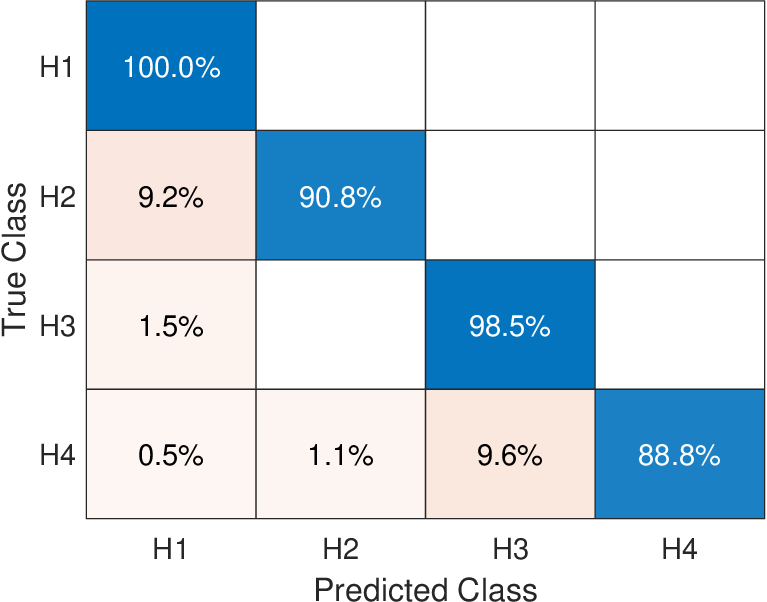}}
    \caption{Confusion matrix for the considered \ac{MOS} rules (for the classes of no symmetry, reflection symmetry, rotation symmetry, and azimuth symmetry), using  $K=49$ looks with $M=2$ and $\bC_t=\bI$. Subplots refer to the \ac{MOS} rules a) AIC, b) BIC, c) GIC, and d) HQC.}
    \label{Simu_2imagesK49}
\end{figure}

To grasp more information on the behavior of the considered selectors, in Figure \ref{Simu_2imagesK49} the confusion matrices are also shown for the same simulation setting as above, but for a larger number of looks, i.e., $K=49$. In this case, also the HQC exceed the $90\%$ threshold value in the detection of the first three structures, approaching this value even in the case of azimuth symmetry. However, it is evident that there is a growth in the performance for all the detection rules regardless of the assumed polarimetric structure. This fact can be explained observing that increasing the number of looks $K$ leads to a more and more accurate estimate of the sample covariance matrix that is at the base of all the estimates obtained from the proposed flip-flop.

Finally, to directly compare the various adopted MOS rules, we present in Table \ref{tab_kappa_coeff} the values of Cohen's kappa coefficient \cite{CohenKappa1960}, $\kappa$, for the same simulation setups as in Figures \ref{Simu_2imagesK25} and \ref{Simu_2imagesK49}. Interestingly, the values of $\kappa$ confirm the trends discussed above. Specifically, an increased number of looks leads to a higher coefficient value for all the proposed selectors. Furthermore, the BIC is confirmed to be the best, achieving a Cohen's kappa coefficient close to 1 in both cases for $K=25$ and $K=49$.

\begin{table}[ht]
\centering
\caption{Cohen's kappa coefficient, $\kappa$, for the considered MOS rules for the simulation setups of Figures \ref{Simu_2imagesK25} and \ref{Simu_2imagesK49}.}
\label{tab_kappa_coeff}
\begin{tabular}{ccccc}
\hline
 & AIC & BIC & GIC & HQC\\ 
\hline
\hline
$K=25$ & 0.83 & 0.95 & 0.94 & 0.89 \\
\hline
$K=49$ & 0.84 & 0.98 & 0.95 & 0.93 \\
\hline
\hline
\end{tabular}
\end{table}

To further corroborate the previous results, in Table \ref{BIC_good_detection_Ct_rho09} the accuracy of each class (i.e., each nominal covariance) is summarized with reference to the BIC rule, having it shown the best performance in the confusion matrices of Figures \ref{Simu_2imagesK25}-\ref{Simu_2imagesK49}. Specifically, the reported values correspond to different number of looks, viz., $K=(6, 9, 25)$, as well as to an exponentially shaped temporal covariance $\bC_t$ with correlation coefficient $\rho=0.9$, whose size $M$ varies from $2$ to $4$. Additionally, for comparison, the results of the classifier designed in \cite{pallotta2016detecting} for a single datum (i.e., $M=1$) are provided to complete the analysis. In agreement with the previous results, the values in the table demonstrate the BIC increasing ability to accurately detect the correct symmetry as the number of available data points grows. For example, with regard to azimuth symmetry and considering $M=2$, it is observed that the accuracy significantly increases from $70.8\%$ when $K=6$ to $92.0\%$ when $K=25$. Analogously, the impact of the number of temporal instances can be also observed analyzing the accuracy as $M$ varies. As an example, with reference to the average accuracy over all symmetries and $K=6$, it can be seen that it increases of an amount more than $6\%$ considering $M=4$ in place of $M=1$. These improved classification results suggest that the overall system performance can be enhanced when a stack of temporal data is used. Finally, it is important to note that the performance is not highly sensitive to temporal correlation (results for different $\rho$ values are omitted for brevity). This, however, can be intuitively explained, as the symmetry is related to the polarimetric part of the covariance matrix, and the Kronecker structure enables the separation of the temporal and spatial contributions.

\begin{table*}[ht!]
    \centering
    \caption{Accuracy ($\%$) using the BIC selection rule for an exponentially shaped $\bC_t$, with $\rho=0.9$.}
    \label{BIC_good_detection_Ct_rho09}
    \begin{tabular}{c|cccc|cccc|cccc} 
    \multicolumn{1}{c}{} & \multicolumn{4}{c}{$K=6$} & \multicolumn{4}{c}{$K=9$} & \multicolumn{4}{c}{$K=25$} \\
    \hline
    \hline
    & $M=1$ &  $M=2$ & $M=3$ & $M=4$ & $M=1$ &  $M=2$ & $M=3$ & $M=4$ & $M=1$ & $M=2$ & $M=3$ & $M=4$\\
    \hline 
    \textbf{No Symmetry} & 99.9\%& 100\% & 100\% & 100\% & 100\% & 100\% & 100\% & 100\% & 100\% & 100\% & 100\% & 100\%\\ \hline 
    \textbf{Reflection symmetry} &  73.4\%&  68.4\% & 70.0\% & 72.1\% & 88.2\% & 80.2\% & 81.23\% & 83.0\% & 98.5\% & 94.6\% & 94.8\% & 94.9\% \\
    \hline 
    \textbf{Rotation symmetry} & 75.2\% & 85.2\% & 87.6\% & 88.0\% & 91.1\% & 94.1\% & 94.9\% & 95.6\% & 99.5\% & 99.6\% & 99.6\% & 99.6\% \\
    \hline 
    \textbf{Azimuth symmetry} & 58.4\% & 70.8\% & 71.7\% & 72.6\% & 74.7\% & 81.0\% & 81.4\% & 81.8\% & 90.6\% & 92.0\% & 92.6\% & 92.6\% \\
    \hline
    \textbf{Average accuracy} & 76.7\% & 81.1\% & 82.3\% & 82.9\% & 88.5\% & 88.8\% & 89.4\% & 90.1\% & 97.1\% & 96.6\% & 96.7\% & 96.8\%\\
    \hline
    \hline
    \end{tabular}
\end{table*}

\begin{figure}[ht!]
    \centering
    \subfigure[proposed method]{
    \includegraphics[width=0.46\columnwidth]{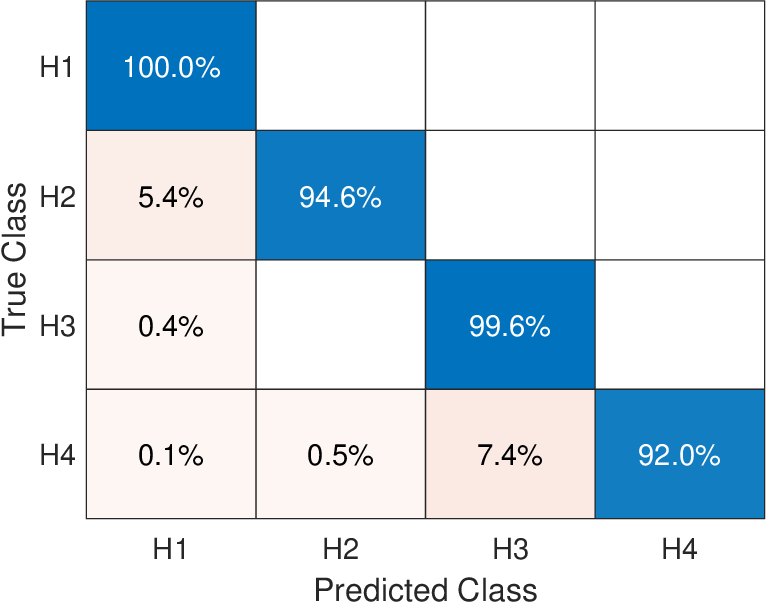}}
    \subfigure[TUSML]{
    \includegraphics[width=0.46\columnwidth]{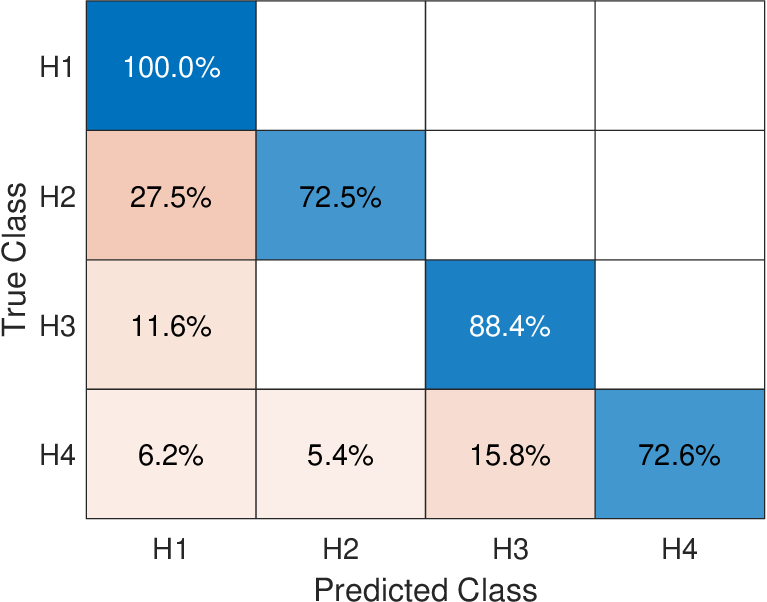}}
    \caption{Confusion matrix for the BIC selection rule (for the classes of no symmetry, reflection symmetry, rotation symmetry, and azimuth symmetry), using  $K=25$ looks with $M=2$ and an exponentially shaped $\bC_t$, with $\rho=0.9$. Subplots refer to a) the proposed structured flip-flop based method, and b) TUSML.}
    \label{Simu_2imagesK25confronto}
\end{figure}

Finally, to complete this quantitative analysis on simulated data, the proposed method using the BIC rule is compared with the corresponding TUSML for the previously described simulation scenario, employing $K=25$ looks and an exponentially shaped temporal covariance $\bC_t$ with a correlation coefficient of $\rho=0.9$ and size $M=2$. The results are still presented in terms of confusion matrices in Figure \ref{Simu_2imagesK25confronto}. A direct comparison of the two matrices clearly shows that, based on the results illustrated above, the proposed framework improves the classification performance compared to the competitor, which does not account for temporal correlation. Focusing on specific values, it is interesting to note that the new method significantly increases the accuracy for the reflection symmetry case with respect to the TUSML, i.e., from $72.5\%$ to $94.6\%$. Analogously, the azimuth symmetry accuracy increases from $72.6\%$ to $92.0\%$. Finally, a comparison of the Cohen's kappa coefficients further quantifies this behavior. In fact, the value achieved by the proposed method is $\kappa = 0.95$, which significantly exceeds that of TUSML, where $\kappa = 0.78$.

\subsection{Assessment on measured data}\label{sub_analysisRealData}

The focus of this section is to evaluate the effectiveness of the proposed covariance symmetry classification scheme for multi-temporal \ac{PolSAR} images. To do this, some experiments on real-recorded \ac{PolSAR} datasets are conducted, concentrating on the use of the BIC selector, since it has shown the best performance on simulated data. Moreover, as for the analyses on simulated data, the proposed flip-flop is run for five iterations and with the estimate of $\bC_t$ initialized to the identity matrix.

\begin{figure}[ht!]
    \centering
    \includegraphics[width=0.95\linewidth]{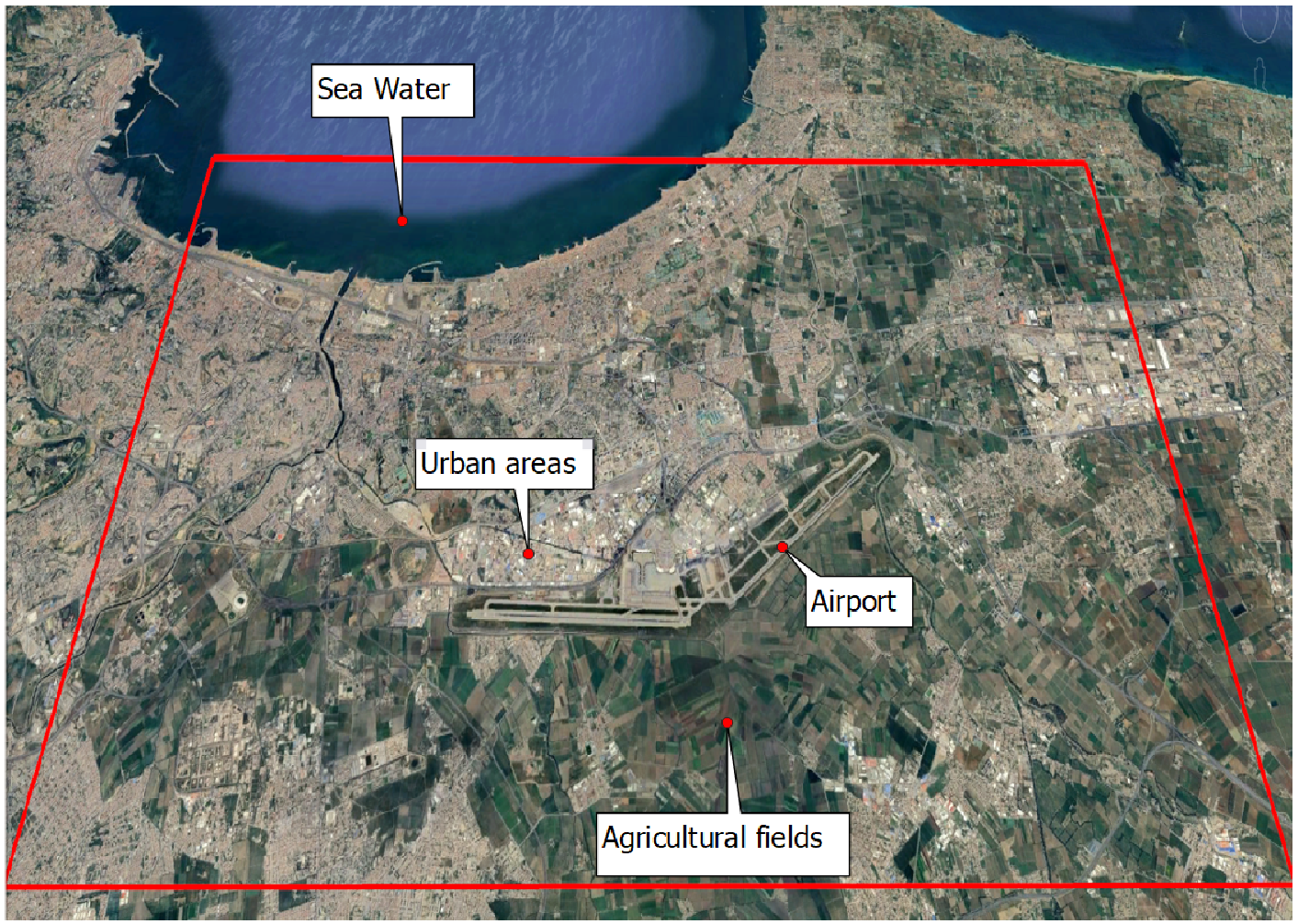}
    \caption{Google Earth$\copyright$ image with a polygon outlining the studied area.}
    \label{fig_googleErth}
\end{figure}

The study is conducted using data collected over an area of Algeria, encompassing regions with diverse surface features and various land uses. This includes a portion of the Mediterranean Sea, coastline, an international airport, urban areas, and agricultural fields. The utilized \ac{PolSAR} data comprise two fully polarized RADARSAT-2 images, acquired on two different dates with an interval of 24 days (i.e., April 11th, 2009, and May 5th, 2009), with the same incidence angles. For our analysis, the selected area of interest is a sub-image of $1750 \times 2500$ pixels (i.e., $L = 1750$ and $C = 2500$). A picture representing the optic image extracted from Google Earth$\copyright$, with a polygon delimiting the area of interest, is drawn in Figure \ref{fig_googleErth}. Finally, the sensor parameters are reported in Table \ref{table_Radarsat_parameters}. 

\begin{table}[ht!]
    \caption{RADARSAT-2 parameters.}
    \centering
	{\begin{tabular}{l c}
	\hline \hline
       Parameter  & Quantity  \\
       \hline
       Wavelength  & 5.5 [cm]  \\
       Sensor altitude  & 798 [km] \\
       Incidence angle  & 38.34 to 39.81 [degrees] \\
       Nominal resolution [range $\times$ azimuth] & $5.2\times 7.6$ [m] \\
       Sampled pixel \& line spacing & $4.73 \times 4.74$ [m] \\
       Look direction & right \\
       Pass & descending \\
       Acquisition mode & fine quad polarization\\
      \hline
\end{tabular}}
    \label{table_Radarsat_parameters}
\end{table}

\begin{figure}[ht!]
    \centering
    \subfigure[]{
    \includegraphics[width=0.9\columnwidth]{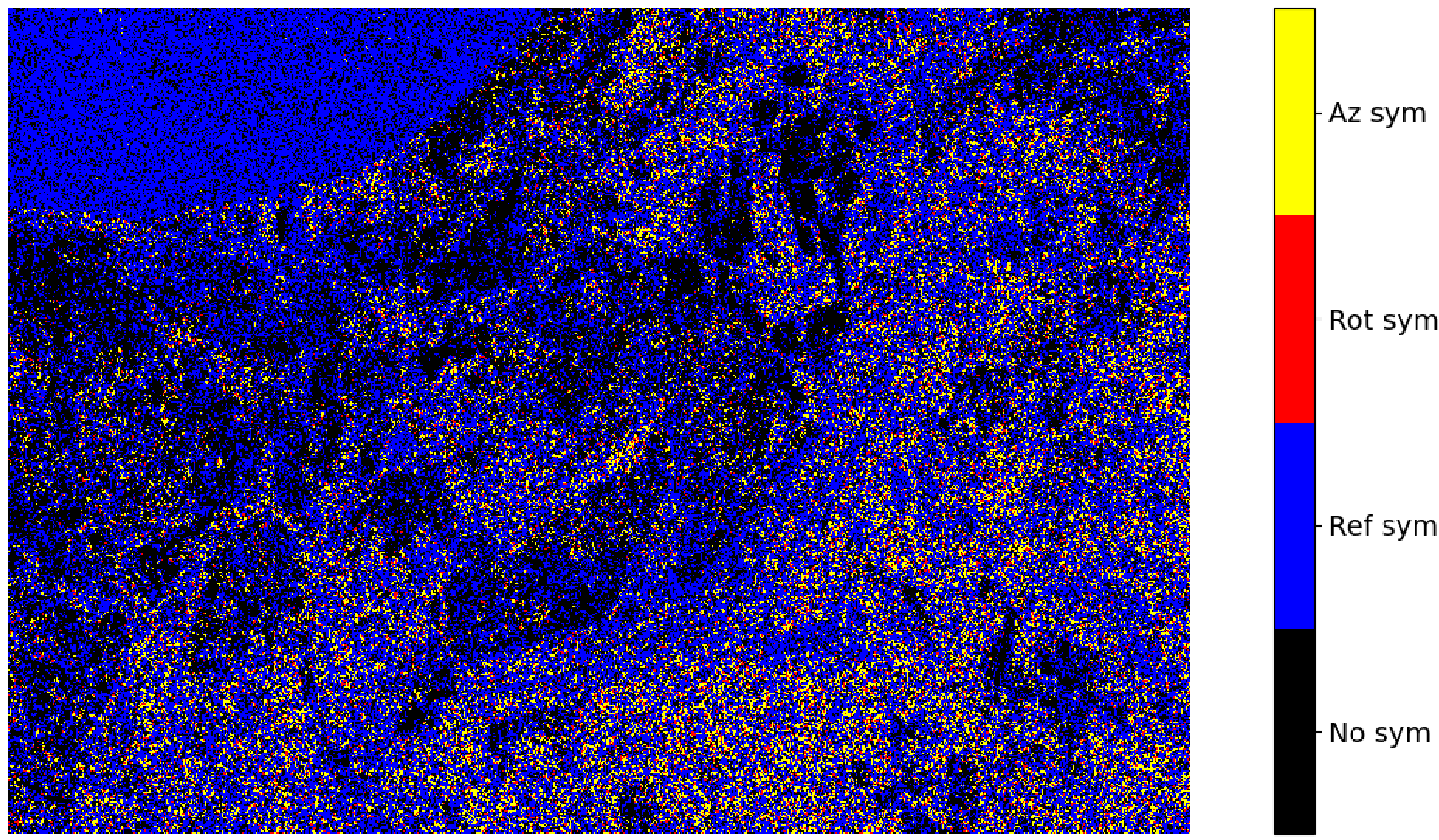}}\\
    \subfigure[]{
    \includegraphics[width=0.9\columnwidth]{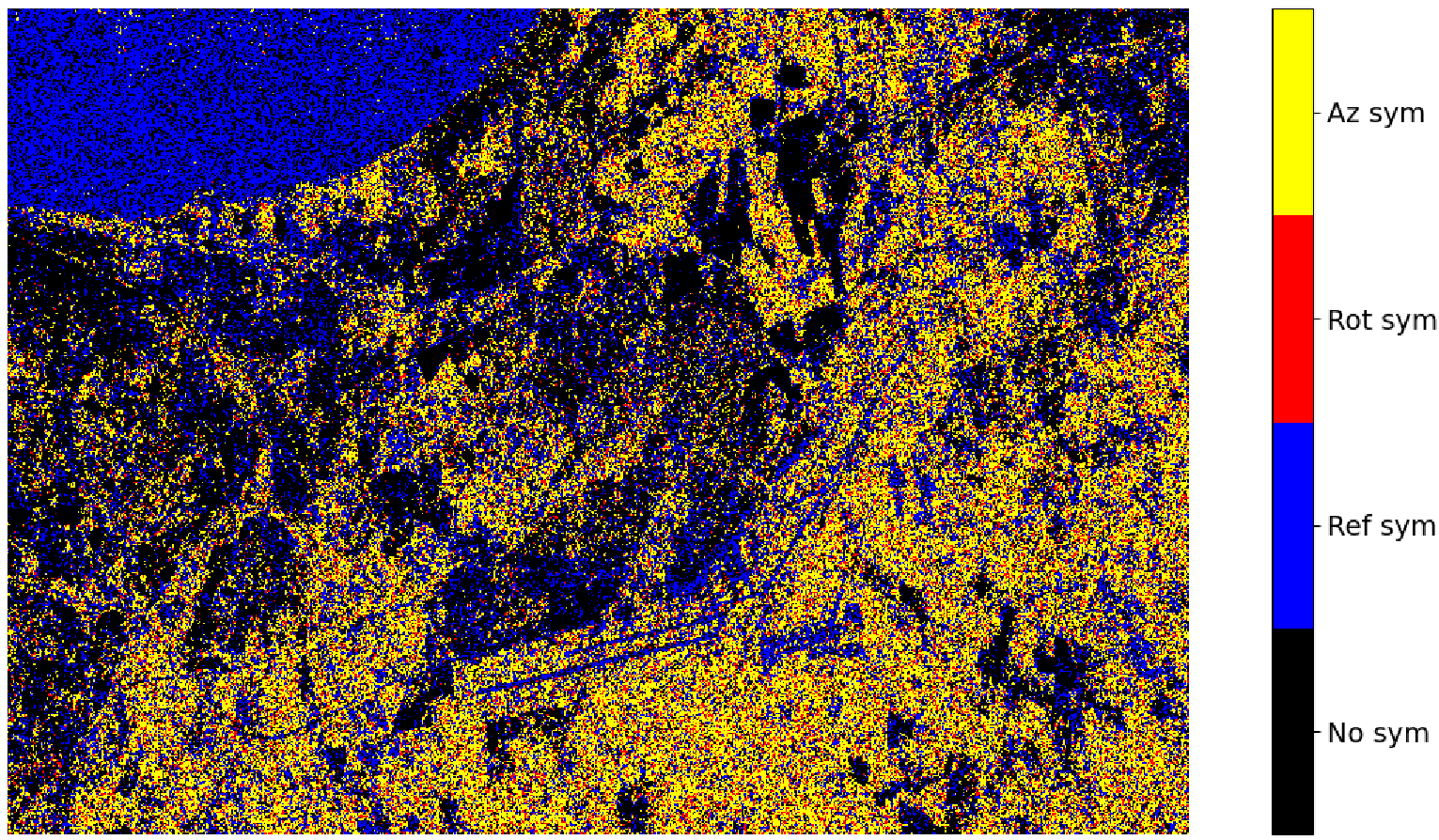}}
    \caption{Symmetries classification using BIC with $K= 25$. Subplots refer to a) algorithm of \cite{pallotta2016detecting} for single image, b) proposed method with $M=2$.}
    \label{BIC_radarsat2_K25}
\end{figure}

In Figure \ref{BIC_radarsat2_K25}, with reference to the RADARSAT-2 images, the symmetries classified through the proposed framework with the BIC rule are plotted. More precisely, the estimations are done using a sliding window of size $5\times 5$, hence with $K=25$. The two subplots refer to a) the method developed in \cite{pallotta2016detecting} for a single image, and b) the proposed algorithm. Moreover, for each pixel of the considered scene, a specific color is used for each hypothesis selected by the test, viz. no symmetry (black), reflection symmetry (blue), rotation symmetry (red), and azimuth symmetry (yellow).

From a first visual inspection of the colored maps, it is evident that the seawater is, as expected, mostly detected as reflection symmetry. Moreover, pixels belonging to the airport runway are also classified as having a reflection symmetry structure. This is not surprising, observing that the runway has a plane structure \cite{Nghiem1992,pallotta2022covariance}. Similarly, also the uncultivated agricultural areas under the airport are classified as being structured with a reflection symmetry. Going further into the details, it can be also ascertained that the vegetation is revealed to be azimuthal symmetry, whereas building are essentially related to absence of symmetry. Finally, the rotation symmetry is detected few times in the entire image, therefore not giving significant information for this specific environment. Now, comparing the results obtained from the algorithm working on the single image with the proposed one exploiting the information embedded in the pair, interesting conjectures can be done. Initially, the most noticeable difference is that areas with some vegetation are much more distinct. The proposed method has classified these areas as having azimuthal symmetry (which aligns with this specific class), whereas the standard method identifies them as having reflection symmetry. Furthermore, in the proposed map, the separation between homogeneous areas appears to be much more evident with sharper boundaries than the counterpart. As a matter of fact, the runway in the proposed case is much better visible than in the single image processing.

Similarly, the coastline appears sharper and better detected. Finally, also building in the image (see the black areas) are better emphasized. These effects are also related to the fact that, using both images as in the proposed methodology, the resulting map appears less noisy with homogeneous areas displayed cleaner. Indeed, using a multi-temporal dataset for the same value of $K$ produces a better classification map than relying on a single image.

\vspace{0.25cm}
\noindent
\textbf{Zone index $\bH/\bar{\balpha}$ decomposition.} Once the polarimetric covariance symmetries have been classified, for each pixel an estimate of the covariance with the detected structure is hence available. Accordingly, a more detailed analysis of their effect on the interpretation of the scene can be performed. Specifically, the classic $H/ \overline{\alpha}$ classification \cite{Cloude1997} is compared with an enhanced version that employs a more accurate estimate, namely, the structured covariance estimates provided by the proposed flip-flop algorithm, instead of the \ac{SCM}\footnote{The structured covariance used in this paper is associated with a polarimetric vector organized as $\by = \left[y_{\text{HH}}, y_{\text{HV}}, y_{\text{VV}}\right]^T$. Differently, to perform the $H/ \overline{\alpha}$ classification, the starting point is the coherence obtained using the Pauli vector \cite{PottierBook}, $\by_P = (1/\sqrt{2})\left[(y_{\text{HH}}+y_{\text{VV}}), (y_{\text{HH}}-y_{\text{VV}}), 2y_{\text{HV}}\right]^T$. For the above reasons, the following transformation must be applied to $\hat{\bC}_p$ in order to derive the proper coherence $\bT_P = \bT \bG \hat{\bC}_p \bG \bT^T$, with 
\vspace{-0.2cm}
\begin{scriptsize}
$$
\bG = \left[\begin{matrix}
1 & 0 & 0\\
0 & \sqrt{2} & 0\\
0 & 0 & 1
\end{matrix}\right].
$$
\end{scriptsize}}. Hence, once the images have been decomposed into the $H/ \overline{\alpha}$, different areas are classified by means of the zone index as proposed in \cite{Cloude1997}, as specified in the following:
\begin{itemize}
\item Class $Z_1$: Double bounce scattering in a high entropy environment;
\item Class $Z_2$: Multiple scattering in a high entropy environment (e.g. forest canopy);
\item Class $Z_3$: Surface scattering in a high entropy environment;
\item Class $Z_4$: Medium entropy multiple scattering;
\item Class $Z_5$: Medium entropy vegetation (dipole) scattering;
\item Class $Z_6$: Medium entropy surface scattering;
\item Class $Z_7$: Low entropy multiple scattering (double or even bounce scattering);
\item Class $Z_8$: Low entropy dipole scattering (strongly correlated mechanisms with a large imbalance in amplitude between HH and VV);
\item Class $Z_9$: Low entropy surface scattering (e.g. Bragg scatter and rough surfaces).
\end{itemize}

\begin{figure}[ht!]
    \centering
    \subfigure[]{
    \includegraphics[width=0.85\columnwidth]{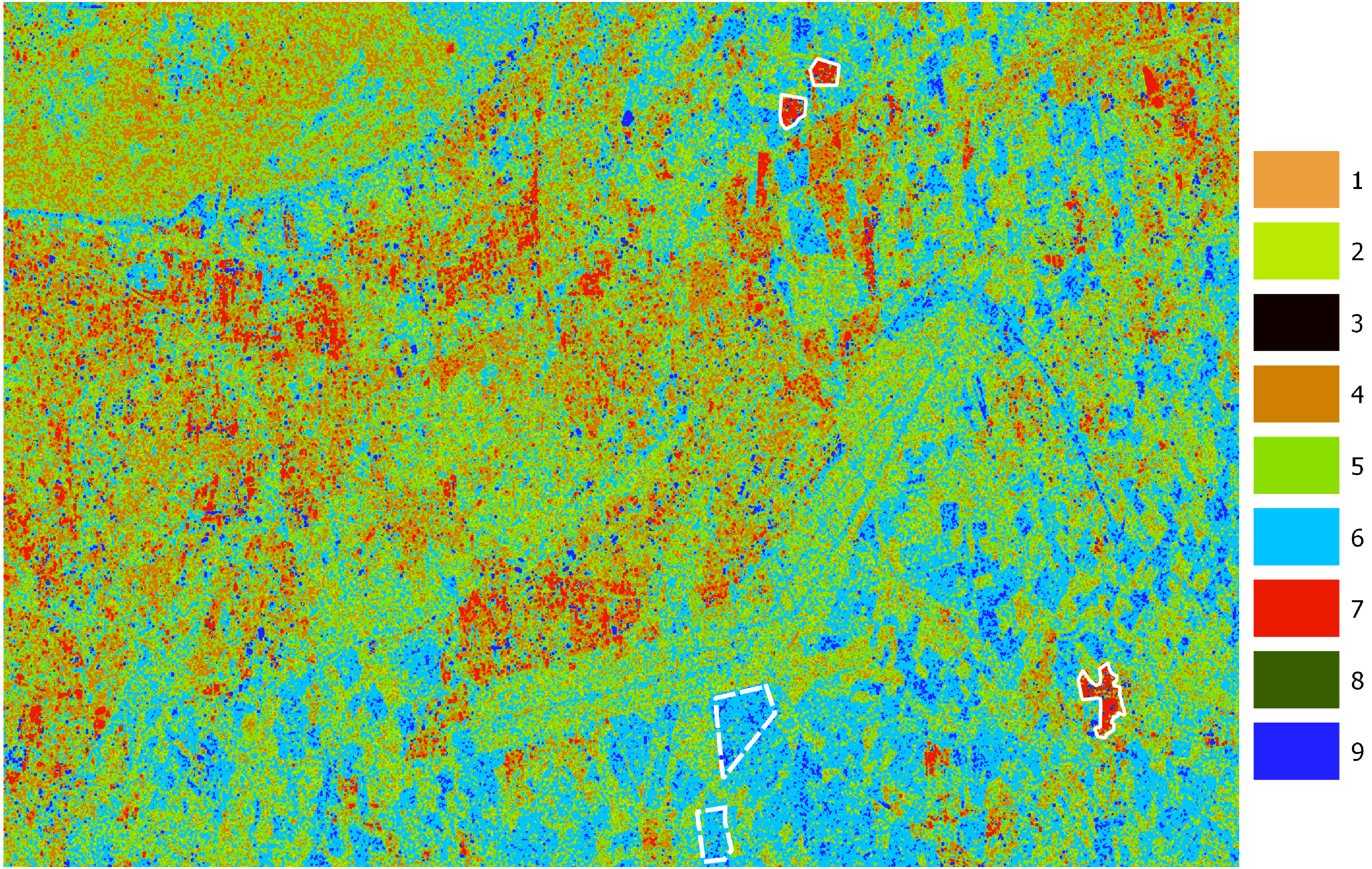}}\\
    \subfigure[]{
    \includegraphics[width=0.85\columnwidth]{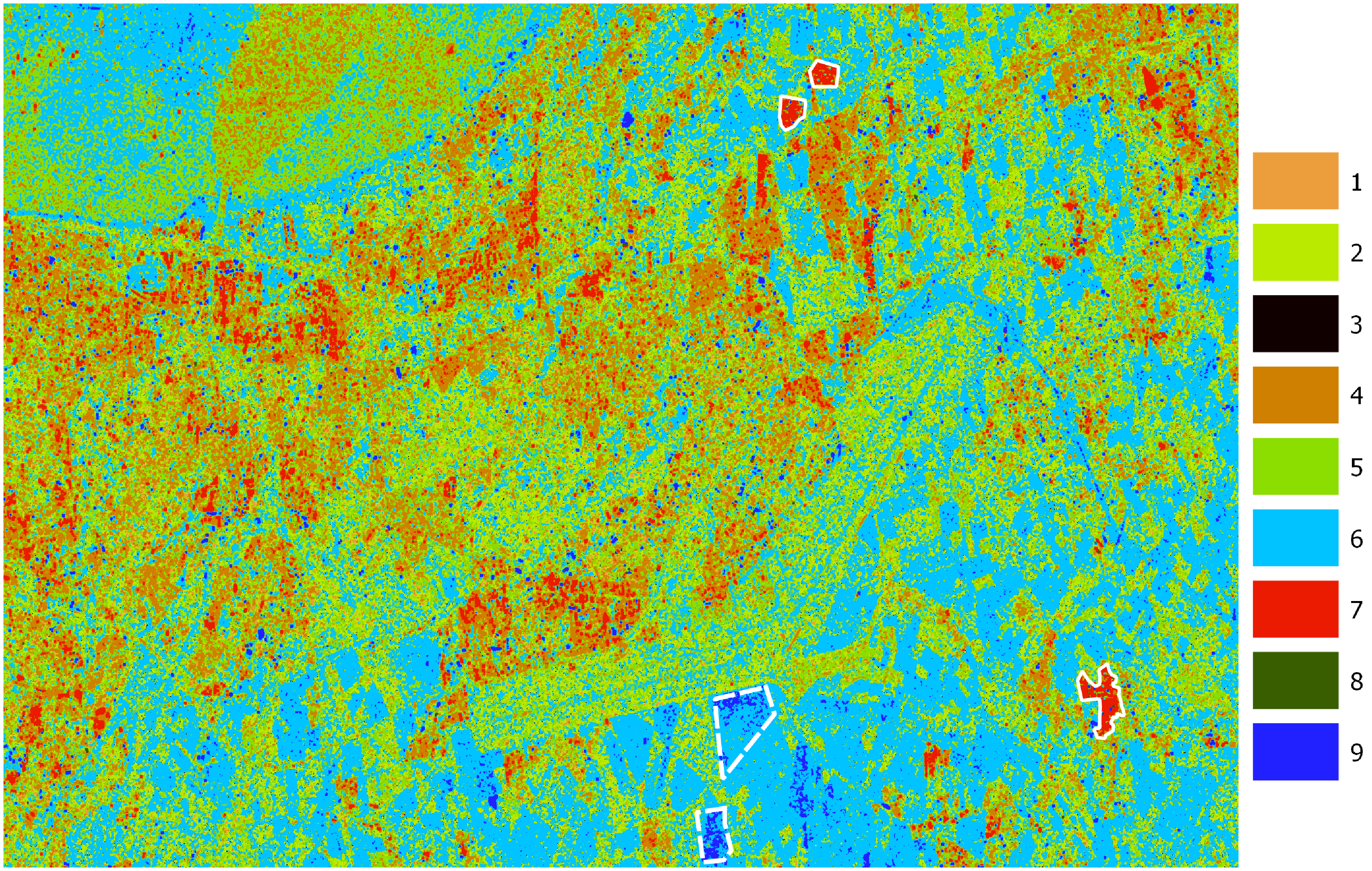}}
    \caption{Zone index $H/\overline{\alpha}$ decomposition of the RADARSAT-2 data using $K=25$ looks. Subplots refer to a) sample coherence matrix, b) reconstructed matrix with the proposed method based on flip-flop.}
    \label{zoneIndex_K25}
\end{figure}

This is reported in Figure \ref{zoneIndex_K25}, where the zone indexes relative to the $H/ \overline{\alpha}$ plane are shown with reference to the classic sample coherence in subplot a) and the proposed structured covariances in subplot b).

Comparing the two zone index maps, it can be observed that the proposed framework improves the classification of buildings as \emph{dihedrals} (class 7), with some unseeded lands being classified as \emph{surface} (class 9) instead of \emph{anisotropic particles} (class 5). Furthermore, marine pollution (see e.g. \cite{hanis2024dominant}) is well illustrated, while it is completely missed in the classic approach. To better quantify these results, two areas of coverage are selected based on the homogeneity and predictability of classes, namely urban areas, and agricultural lands. These are indicated in Figure \ref{zoneIndex_K25}, where dense urban areas refer to the three patches delineated by solid lines, whereas the agricultural lands are indicated with the dashed lines. For each environment mentioned above, the percentage of pixels belonging to the classes detected according to the H/$\bar{\alpha}$ plane is reported in Table \ref{tab_histogramZindex}. To go into detail, it can be seen that for urban areas, the percentage of pixels associated with the class 7 describing multiple scattering (as expected for this environment) increases from $60.01\%$ with the classic (sample covariance based) H/$\bar{\alpha}$ plane to $68.2\%$ with the proposed approach. Regarding agricultural land zones, instead, it is clearly observed that the proposed method increases the percentage of pixels classified as $Z_9$ (i.e., surface), from $13.56\%$ with the classic method to $26.52\%$. This is not surprising, since a lower entropy indicates more homogeneous targets, as expected for the agricultural fields. Finally, being class $Z_3$ a \emph{nonfeasible region} in the $H/\bar{\alpha}$ plane \cite{PottierBook}, it is desidered a number as small as possible of entries classified as belonging to this class (ideally empty). However, in some cases for the agricultural lands, a few pixels could be misclassified as $Z_3$ due to factors such as numerical errors or convergence issues related to the decomposition process. Interestingly, the proposed approach allows reducing this percentage from $0.28\%$ of the standard method to $0.13\%$.

\begin{table*}[ht]
\centering
\caption{Inferred scattering mechanism over urban area and agricultural fields.}
\label{tab_histogramZindex}
\begin{tabular}{ccccccccccc}
\textbf{area} & \textbf{method} & \boldmath{$Z_{1}$} & \boldmath{$Z_{2}$} & \boldmath{$Z_{3}$} & \boldmath{$Z_{4}$} & \boldmath{$Z_{5}$} & \boldmath{$Z_{6}$} & \boldmath{$Z_{7}$} & \boldmath{$Z_{8}$} & \boldmath{$Z_{9}$}\\ 
\hline
\hline
\multirow{2}{*}{urban} & classic & $0.16\%$ & $0.30 \%$ & $0 \%$ & $14.38\%$ & $11.84\%$ &  $4.12\%$ & $60.01\%$ & $3.60\%$ &  $5.59\%$ \\
& proposed & $0.22\%$ & $0.55\%$ & $0 \%$ & $17.45\%$ &  $7.88\%$ &  $1.99\%$ & $\mathbf{68.2 \%}$ & $1.35\%$ &  $2.36\%$ \\
\hline
\multirow{2}{*}{agricultural} & classic & $0\%$ & $0.65\%$ &   $0.28\%$ &   $0\%$ &    $7.78\%$ &   $77.65\%$ &   $0.08\%$ & $ 0\%$ &  $13.56\%$ \\
& proposed & $0.01\%$ & $0.84\%$ & $\mathbf{0.13 \%}$ & $0.03 \%$ & $1.01\%$ & $71.44\%$ & $0.01\%$ & $0.01\%$ & $\mathbf{26.52\%}$ \\
\hline
\end{tabular}
\end{table*}

\begin{figure}[ht!]
    \centering
    \subfigure[]{
    \includegraphics[width=0.85\columnwidth]{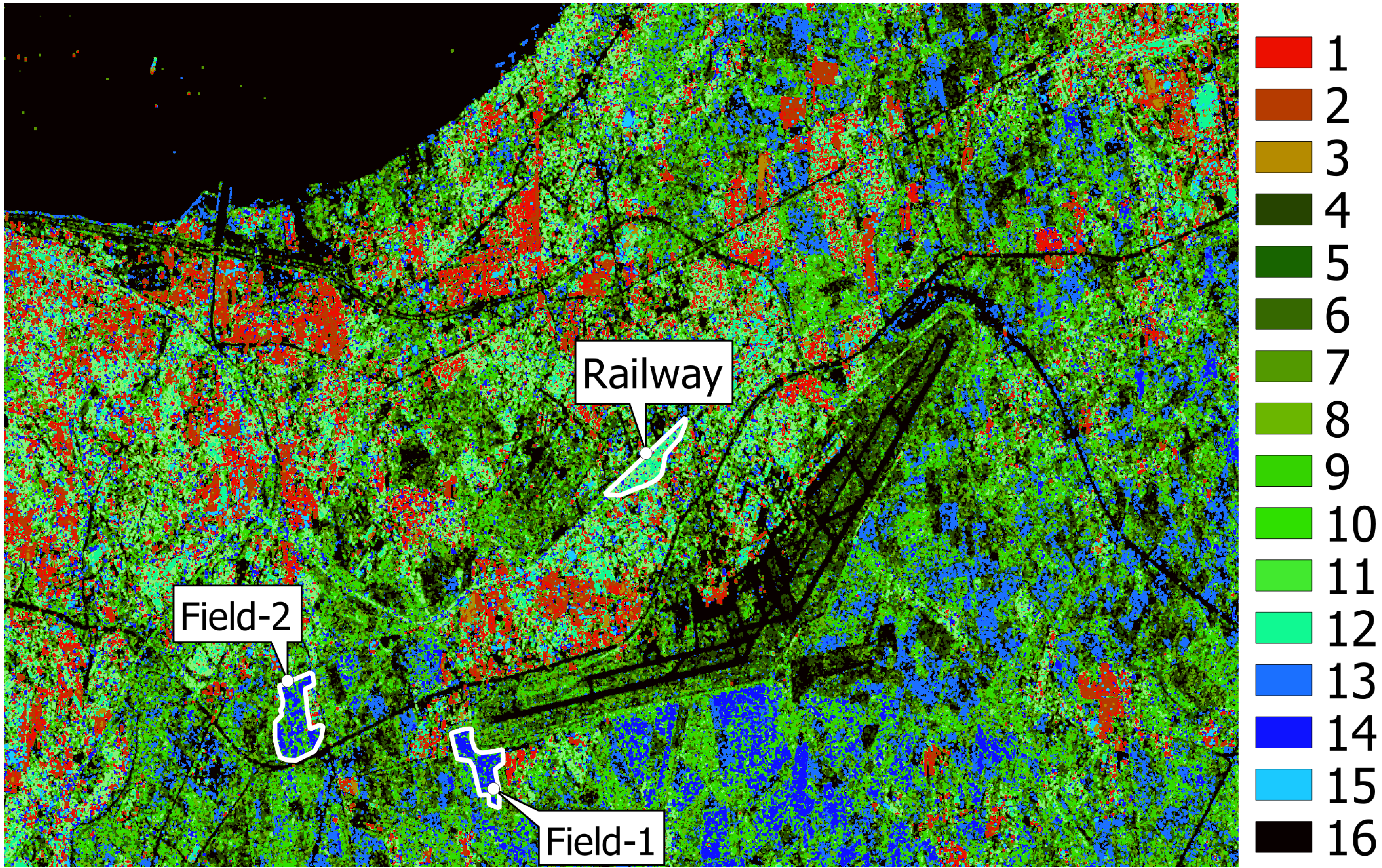}}\\
    \subfigure[]{
    \includegraphics[width=0.85\columnwidth]{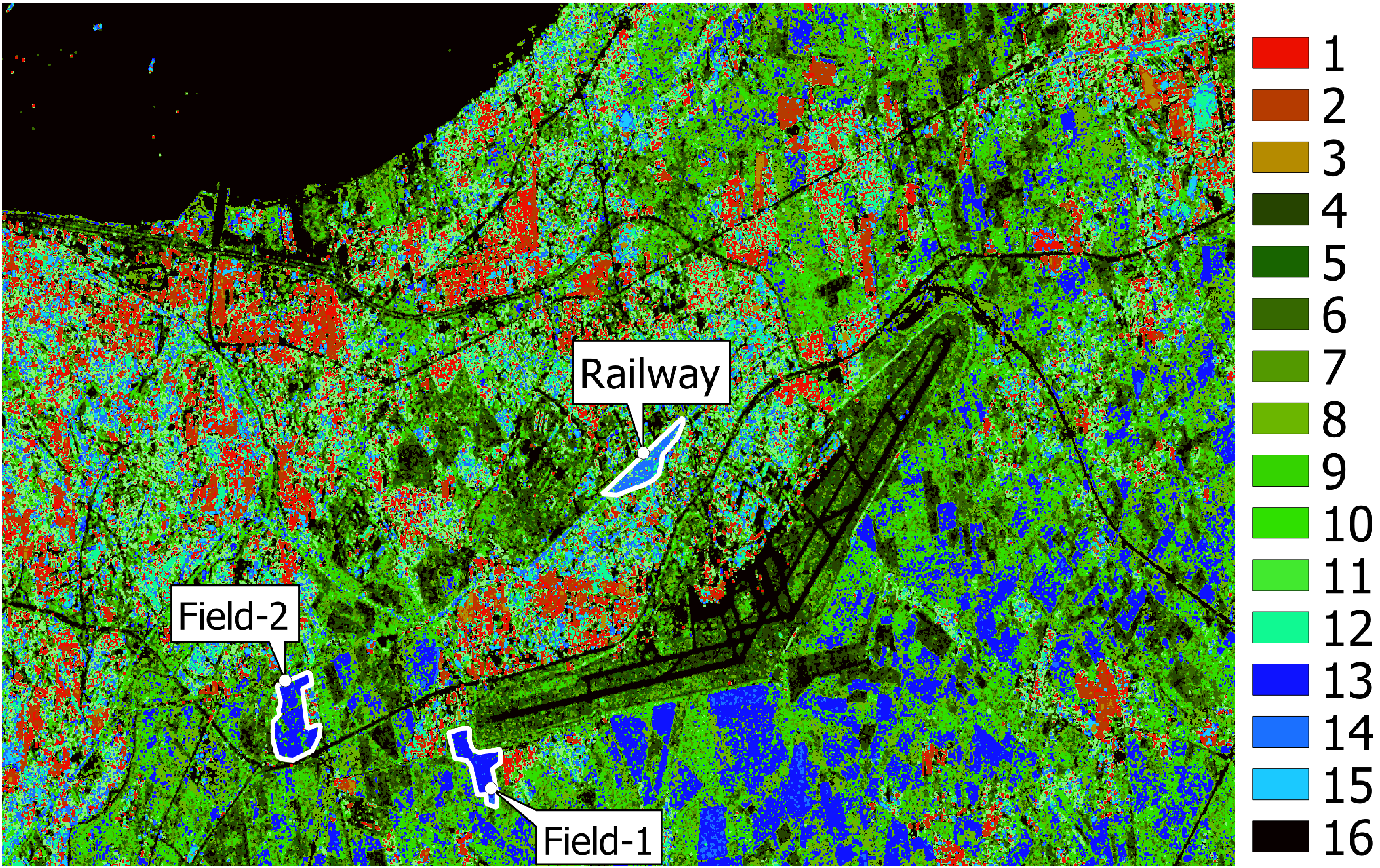}}
    \caption{Freeman-Durden Wishart classification of the RADARSAT-2 data using  $K=25$ looks. Subplots refer to a) sample coherence matrix (i.e., classical implementation), b) reconstructed matrix with the proposed method based on flip-flop.}
    \label{freemanDurdenDecom_K25}
\end{figure}

\begin{figure}[ht]
    \centering
    \subfigure[]{\includegraphics[width=0.85\linewidth]{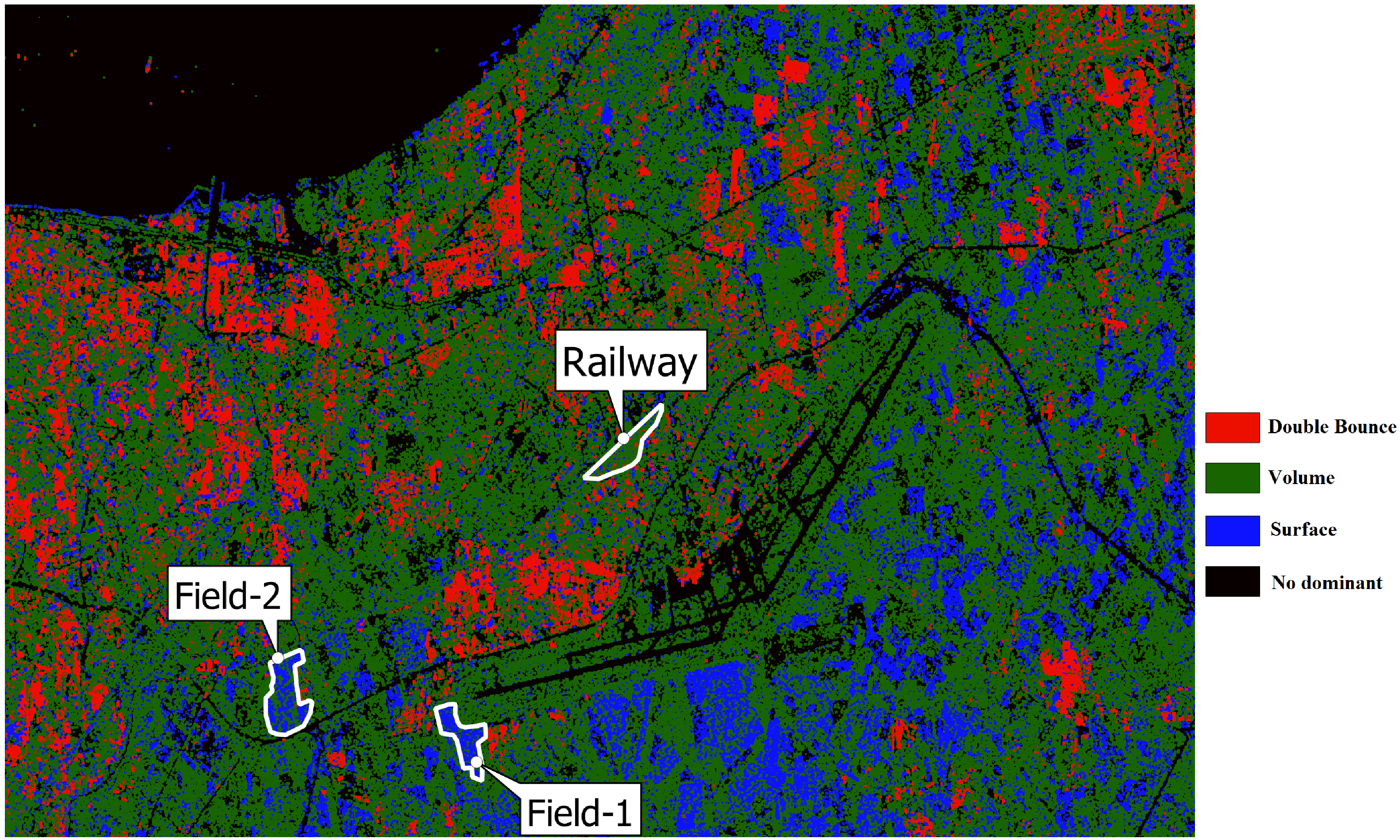}}
    \subfigure[]{\includegraphics[width=0.85\linewidth]{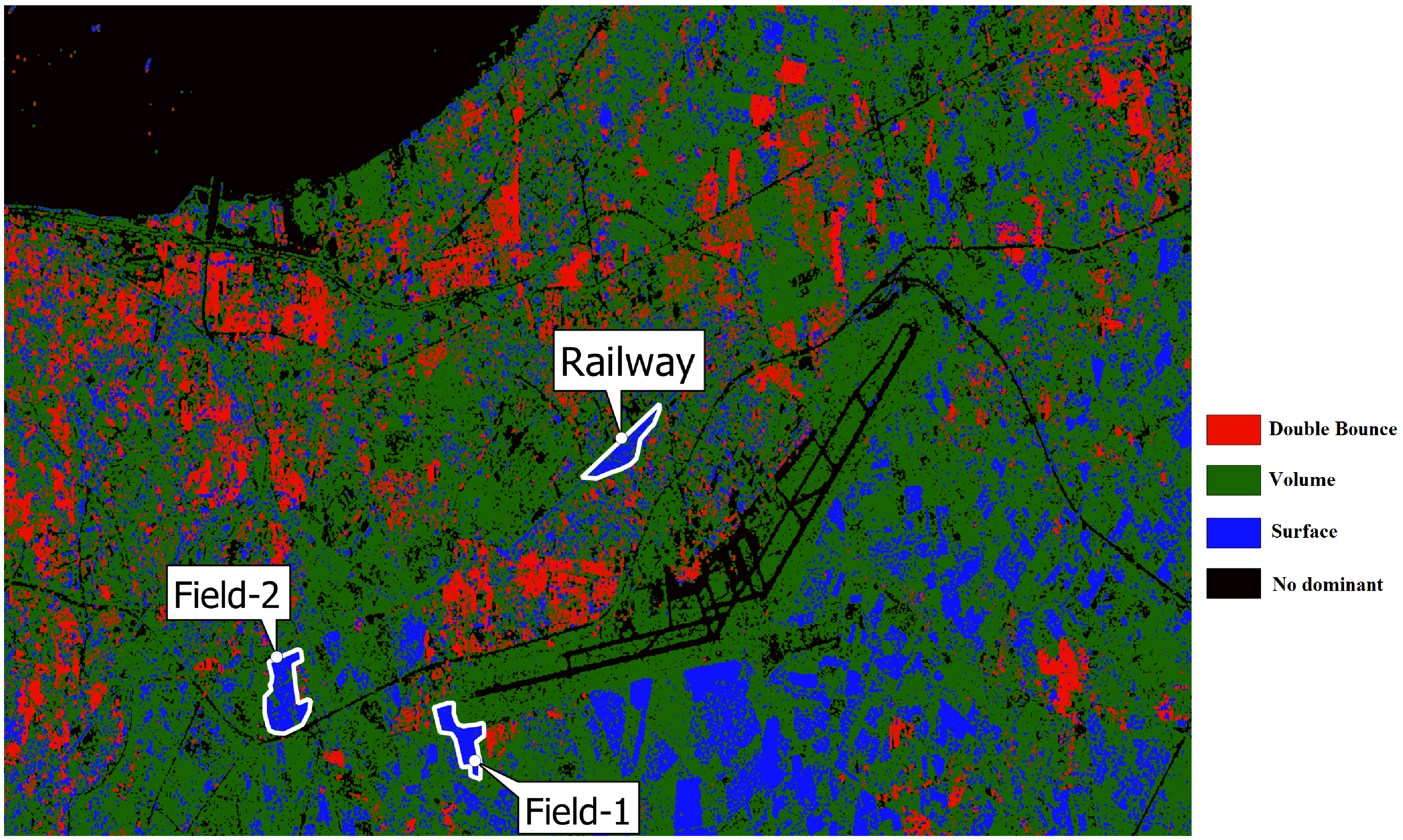}}
   \caption{Inferred dominant scattering mechanisms over the RADARSAT-2 scene using different coherence matrix estimates. Subplots refer to a) sample coherence matrix (i.e., classical implementation), and b) reconstructed matrix with the proposed method based on flip-flop.}
    \label{Fig_dominant_scattering_detected}
\end{figure}

\vspace{0.25cm}
\noindent
\textbf{Freeman-Durden Wishart classification.} The last analysis consists in performing the Freeman-Durden Wishart classification as proposed in \cite{1288367} to further show the impact of a more accurate covariance estimate in emphasizing details within the image. This method categorizes the data into three main types based on the dominant scattering mechanisms: 
\begin{enumerate}
    \item surface scattering, 
    \item volume scattering,
    \item double bounce.
\end{enumerate}
The above categories are further subdivided into 16 subclasses \cite{1288367}. Moreover, a fourth category is also designated for mixed scattering pixels that do not clearly fit into any of the three primary scattering mechanisms (it is denoted as {\em no dominant}). The results of the classification process are illustrated in Figure \ref{freemanDurdenDecom_K25}, while the subclasses grouped according to the dominant mechanism are presented in Figure \ref{Fig_dominant_scattering_detected}. The procedure is done again using the estimated $\bC_p$ relying on the proposed BIC selection rule, in comparison with the classic Freeman-Durden Wishart classification initialized by the sample coherence.

To quantify the results, the analysis of the classification process based on the two coherence estimates are conducted over three distinct zones. The zones, highlighted in Figure \ref{freemanDurdenDecom_K25} and labelled as {\em Field 1}, {\em Field 2}, and {\em Railway}, would normally exhibit a dominant surface scattering mechanism. As a matter of fact, {\em Field 1} and {\em Field 2} correspond to two bare-soil fields, whereas {\em Railway} contains a railway. In Table \ref{tab_1} the percentage of pixels classified according to the three main Freeman-Durden classes is presented for the three analyzed zones, comparing the different methods. As shown in the tables, the proposed flip-flop based method significantly increases the percentage of pixels classified as surface scattering across all three zones. Specifically, for {\em Field 1}, the percentage of correctly classified pixels reaches $91.65\%$, compared to $68.23\%$ with the classic method. A similar behavior is observed for {\em Field 2}. Interestingly, for the {\em Railway} zone, the new method classifies $65.94\%$ of pixels as surface scattering, while reducing the double-bounce classification to just $0.30\%$.

\begin{table}[ht]
\centering
\caption{Inferred scattering mechanism over \emph{Field-1}, \emph{Field-2}, and \emph{Railway} for the RADARSAT-2 image.}
\label{tab_1}
\begin{tabular}{cccccc}
\textbf{zone} & \textbf{method} & \textbf{surface} & \textbf{volume} & \textbf{double-} & \textbf{no-dominant} \\ 
& & & & \textbf{bounce} & \\
\hline
\hline
\multirow{2}{*}{\emph{Field-1}} & classic & $68.23\%$ & $31.12\%$ & $0 \%$ & $0.65 \%$\\ 
& proposed & $91.65\%$ & $8.19\%$ & $0 \%$ & $0.16 \%$\\ 
\hline
\multirow{2}{*}{\emph{Field-2}} & classic & $53.91\%$ & $44.94\%$ & $0.14\%$ & $1.01\%$ \\ 
& proposed & $77.10\%$ & $22.12\%$ & $0.13\%$ & $0.65\%$\\ 
\hline
\multirow{2}{*}{\emph{Railway}} & classic & $10.59\%$ & $81.30\%$ & $5.32\%$ & $2.79\%$\\ 
 & proposed & $65.94\%$ & $32.43\%$ & $0.30\%$ & $1.33 \%$\\ 
\hline
\end{tabular}
\end{table}

\section{Conclusions}\label{sec_conclusion}

In this paper, a new method for identifying covariance symmetries in multipass \ac{PolSAR} images has been developed and examined. Precisely, the proposed algorithm exploits the temporal correlation available from repeated acquisitions to enhance the estimation of the polarimetric structure. Specifically, the covariance matrix of the data has been modelled as the Kronecker product between temporal (unstructured) and polarimetric (having some specific structures) covariances. The solution of the resulting constrained \ac{ML} estimation problem has been obtained through the iterative alternating optimization flip-flop algorithm. Finally, some \ac{MOS} criteria have been synthesized in order to properly select the structure of the polarimetric covariance matrix that provides the best representation of the scattering mechanism for a given pixel. The effectiveness of the proposed methodology has been proven by extensive simulations, demonstrating its advantages with respect to the competitor which does not exploit temporal correlations. Specifically, the proposed method is capable of reaching an accuracy of $94.6\%$ for the reflection symmetry class and $92.0\%$ for the azimuth class, in a typical simulated scenario, whereas the competitor obtain $72.5\%$ and $72.6\%$ for two classes, respectively, under the same simulation conditions. As a matter of fact, the Cohen's kappa coefficient achieved by the proposed method is $\kappa = 0.95$, which significantly exceeds that of TUSML, where $\kappa = 0.78$. Moreover, tests conducted on measured X-band SAR data of RADARSAT-2 have shown its capabilities to enhance the image interpretation when it is plugged in widely used polarimetric decompositions. In particular, regarding the Freeman-Durden Wishart classification, tests have shown that the devised flip-flop based method significantly increases the percentage of correctly classified pixels. In fact, in zones dominated by surface scattering, it raises the percentage of correctly classified pixels from $68.23\%$ with the classic method to $91.65\%$. Moreover, it also reduces the erroneously classified double-bounce pixels from $5.32\%$ to just $0.30\%$. Interestingly, the analyses on the real SAR data have demonstrated the flexibility of the designed approach to be integrated into operational contexts and algorithms.

Future researches may consider the extension of this methodology to account for the presence of spatial heterogeneity within the data. Moreover, it could be of interest to consider some structures also on the temporal covariance (such as the Toeplitz structure \cite{du2020toeplitz, aubry2021new, aubry2022ATOM}) to account for the statistical stationarity of the data acquisition mechanism.

\section*{Acknowledgment}

The authors would like to thank the Canadian Space Agency for supplying, as part of the SOAR program under the $\#$1670 project, the polarimetric RADARSAT-2 data used in this paper.

\bibliographystyle{IEEEbib}
\bibliography{biblio}

\end{document}